\documentclass[10pt]{article}

\usepackage{amsmath}
\usepackage{amsthm}
\usepackage{amssymb}
\usepackage{graphicx}
\usepackage{appendix}

\input{Formatting}

\newcommand{\lb}{\left(}
\newcommand{\rb}{\right)}

\newtheorem{thm}{Theorem}
\newtheorem{lemma}[thm]{Lemma}
\newtheorem{prop}[thm]{Proposition}

\renewcommand{\a}{\alpha}
\newcommand{\eps}{\varepsilon}

\newcommand{\dx}{\partial_x}
\newcommand{\Dx}{|\!\dx\!|}
\newcommand{\dt}{\partial_t}
\newcommand{\dc}{\partial_c}
\newcommand{\p}{\partial}
\newcommand{\dd}[1]{\frac{d}{d #1}}
\newcommand{\RegSympOp}{\p_\a^{-1}}

\newcommand{\SympOp}{{\mathcal K}}
\newcommand{\SympOpInv}{\dx}

\newcommand{\AnisoRest}{K_{Q\a}}
\newcommand{\AnisoRestNormlzdInv}{{\Omega_{Q\a}^{-1}}}
\newcommand{\AnisoRestNormlzd}{{\Omega_{Q\a}}}
\newcommand{\AnisoRegSympOp}{{{\mathcal K}_{Q\a }}}
\newcommand{\intR}[1]{\int_{-\infty}^\infty #1\, dx}

\newcommand{\NullProj}{P_{Q}}
\newcommand{\NullProjBar}{\bar{P}_{Q}}
\newcommand{\TMProj}{{\cal P}_{Q}}
\newcommand{\Lca}{\mathcal{L}_{Q}}

\newcommand{\comm}[2]{\left[#1,\,#2\right]}

\newcommand{\Null}[1]{\mbox{\rm Null}\ #1}

\newcommand{\Span}[1]{\mbox{\rm Span}\ \left\{#1\right\}}

\newcommand{\HamiltonianWithPotential}{H_\B}
\newcommand{\HamiltonianWithoutPotential}{H_{\B=0}}

\newcommand{\ActionAtSoliton}{\Lambda_{c a}}
\newcommand{\LyapunovFunctional}{M_c}

\renewcommand{\O}[1]{\mbox{\rm O}\left( #1\right)}

\newcommand{\Qc}{Q_c}
\newcommand{\Qca}{Q_{c a}}

\newcommand{\vco}{\zeta^n_c}

\newcommand{\va}{\zeta^{tr}_{c a}}
\newcommand{\vc}{\zeta^n_{c a}}

\newcommand{\e}{\xi}

\newcommand{\etabot}{{\eta_\bot}}
\newcommand{\emin}{{\e_b}}
\newcommand{\ep}{{\e_g}}

\newcommand{\B}{b}
\newcommand{\DB}{\delta\! \B}
\newcommand{\RB}{\delta^2\!\B}
\newcommand{\ev}{\epsilon_x}
\newcommand{\av}{\epsilon_a}
\newcommand{\tv}{\epsilon_t}

\newcommand{\NpA}[1]{N'(#1)}
\newcommand{\NA}[1]{N(#1)} 

\newcommand{\R}{{\mathbb R}}
\newcommand{\Rp}{\R_+}

\newcommand{\C}{{\mathbb C}}
\newcommand{\Lp}[1]{L^{#1}(\R)}
\newcommand{\Hs}[1]{H^{#1}(\R)}

\newcommand{\TcaM}{T_{\Qca}M_s}

\newcommand{\LpNorm}[2]{\left\|#2\right\|_{L^{#1}}}
\newcommand{\LpXNorm}[2]{\left\|#2\right\|_{L^{#1}_X}}
\newcommand{\LpTNorm}[2]{\left\|#2\right\|_{L^{#1}_T}}

\newcommand{\InfI}[1]{\left\lfloor #1\right\rfloor}
\newcommand{\HsNorm}[2]{\left\|#2\right\|_{H^{#1}}}
\newcommand{\HsXNorm}[2]{\left\|#2\right\|_{H_X^{#1}}}
\newcommand{\HTNorm}[1]{\left\|#1\right\|_{T}}

\newcommand{\SupNorm}[1]{\left\|#1\right\|_\infty}
\newcommand{\ip}[2]{\left\langle #1,#2\right\rangle}
\newcommand{\LTLXNorm}[3]{\left\|#3\right\|_{L_T^{#1} L_X^{#2}}}
\newcommand{\LTWXNorm}[4]{\left\|#4\right\|_{L_T^{#1} W_X^{#2,#3}}}
\newcommand{\LTHXNorm}[3]{\left\|#3\right\|_{L_T^{#1} H_X^{#2}}}
\newcommand{\LXLTNorm}[3]{\left\|#3\right\|_{L_X^{#1} L_T^{#2}}}

\title{Long-Time Dynamics of KdV Solitary Waves over a Variable Bottom\thanks{This paper is part of the first author's Ph.D. thesis.}}
\author{S.I. Dejak$^{1\ddagger}$\thanks{Supported by NSERC under grant NA7901 and Ontario Graduate Scholarships.} and I.M. Sigal$^{1}$\thanks{Supported by NSF under grant DMS-0400526.}\vspace{5mm}\\
\small $^1$University of Notre Dame, Notre Dame, U.S.A.\\\small
$^1$University of Toronto, Toronto, Canada}
\begin{document}
\maketitle

{\abstract We study the variable bottom generalized Korteweg-de
Vries (bKdV) equation $\dt u=-\dx\lb\dx^2 u+f(u)-\B(t,x)u\rb$,
where $f$ is a nonlinearity and $\B$ is a small, bounded and
slowly varying function related to the varying depth of a channel
of water. Many variable coefficient KdV-type equations, including
the variable coefficient, variable bottom KdV equation, can be
rescaled into the bKdV. We study the long time behaviour of
solutions with initial conditions close to a stable, $\B=0$
solitary wave.  We prove that for long time intervals, such
solutions have the form of the solitary wave, whose centre and
scale evolve according to a certain dynamical law involving the
function $\B(t,x)$, plus an $\Hs{1}$-small fluctuation.}

\fixNumberingInArticle
\section{Introduction}
We study the long time behaviour of solutions to a class of
Korteweg-de Vries type equations, which we call the variable
bottom generalized KdV equation (bKdV).  These equations are of
the form
\begin{align}
 \dt u=-\dx\lb\dx^2 u+f(u)-\B(t,x)u\rb,
\label{Eqn:KdvGeneralizedWithPotential}
\end{align}
where $f$ is a nonlinearity and $\B(t,x)$ is a real function. When
$f(x)=x^2$, the bKdV is related to an equation for the bottom of
the channel appearing in the derivation of the KdV from shallow
water wave theory. Examples of possible choices for the
nonlinearity are $f(u)=u^2$, the Kortweg-de Vries (KdV) from
shallow water wave theory; $f(u)=u^3$, the modified KdV (mKdV)
from plasma physics; and $f(u)=u^p$, the generalized power
nonlinearity KdV (gKdV). When $\B=0$,
\eqref{Eqn:KdvGeneralizedWithPotential} reduces to the generalized
Korteweg-de Vries equation (GKdV)
\begin{equation}
\dt u=-\dx(\dx^2 u+f(u)). \label{Eqn:GKdV}
\end{equation}

The KdV is obtain by unidirectionalizing the small amplitude, long
wave/shallow water limit of the two dimensional water wave system
with a constant bottom.  The first such derivation was given by
Korteweg and de Vries \cite{KoDe1895} over a century ago in an
attempt to explain the existence of solitary waves of permanent
form in a shallow channel.  Numerous authors have improved the
formal derivation using either asymptotic expansions \cite{Wi1974}
or Hamiltonian methods \cite{CrGr1994,CrSu2000}.  Schneider and
Wayne \cite{ScWa2000} have given a rigorous proof of the validity
of the KdV in approximating the water wave system in the KdV
regime over time intervals of $\O{1}$.  The KdV also appears in
algebraic geometry. A nice survey of the KdV and it's relation to
algebraic geometry is given by Arbarello \cite{Ar2000}.

A remarkable property of the GKdV is the existence of spatially
localized solitary (or travelling) waves, i.e. solutions of the
form $u=\Qc(x-a-c t)$, where $a\in\R$ and $c$ in some interval
$I$. When $f(u)=u^p$ and $p\ge 2$, solitary waves are explicitly
computed to be
\begin{equation*}
\Qc(x)=c^\frac{1}{p-1}Q(c^\frac{1}{2}(x-c t)),
\end{equation*}
where
\begin{equation*}
Q(x)=\lb\frac{p+1}{2}\rb^\frac{1}{p-1} \lb\cosh\lb\frac{p-1}{2}
x\rb\rb^2.
\end{equation*}
It is generally believed that an arbitrary, say $\Hs{1}$, solution
to equation \eqref{Eqn:GKdV} eventually breaks up into a
collection of solitary waves and radiation.  A discussion of this
phenomenon for the generalized KdV appears in Bona \cite{BoSo94}.
 For the general, but integrable, case see Deift and Zhou \cite{DeZh93}.

As channels with constant bottom do not exist, it is of interest
to know how solutions initially close to a solitary wave behave as
the wave propagates over channels with a variable bottom.
Derivations of KdV-type equations when the bottom varies slowly
have been presented by numerous authors.  See, for example
\cite{CrGu2004,GrPu1993,Jo1973,Mi1979,YoLi1994}. The resulting
equations are the KdV with variable coefficients depending on the
variable bottom.  These derivations are non-rigorous and agree to
the leading order in the bottom length, i.e. in order $\sup |\dx
h(x)|$.  We assume a depth $h(x)$ of $\O{1}$ with length scale
$l_h$.  Consider solutions of the water wave system with
wavelength scale $l_\lambda$, wave amplitude scale $l_\eta$ and
fluid velocity scale $l_u$.  Then, if these scales are related as
$l_\lambda=\O{\eps^\frac{1}{2}}$, $l_\eta=\O{\eps}$,
$l_u=\O{\eps}$ and $l_h=o(\eps^\frac{3}{2})$, the leading equation
for the wave amplitude, after an additional rescaling of the time
variable, is (see equation (74) in \cite{GrPu1993})
\begin{equation*}
\dt\eta=-\Gamma\lb\frac{\eta}{\epsilon}+\frac{3}{4
h}\eta^2+\frac{1}{6}\dx\lb h^2\dx\eta\rb\rb.
\end{equation*}
Here $\Gamma$ is the anti-symmetric operator
$\Gamma:=\frac{1}{2}\lb c(x)\dx+\dx c(x)\rb$, with $c(x)=\sqrt{g
h(x)}$ ($g$ is gravitational acceleration), and
$\epsilon\eta(\epsilon^{-\frac{1}{2}}x,\epsilon^{-\frac{3}{2}}t)$
is the surface elevation measured from the flat interface $y=0$.
We assume $h=h_0+h_1$ with $|h_1|<<h_0$ and $h_0$ a nonzero
constant.  Dropping terms of $\O{\dx h}$ in the above equation for
$\eta$, and changing variables as $\eta(x,t)=v(y,t)$, where
$y=x-\frac{c_0}{\epsilon}t$ and $c_0=\sqrt{g h_0}$, leads to an
equation for $v$:
\begin{equation*}
\dt v=-\p_y\lb \frac{c-c_0}{\epsilon}v+\frac{3}{4
h}cv^2+\frac{1}{6}c h^2\p_y^2 v \rb.
\end{equation*}
To order $\O{h_1}$, solutions of this equation and solutions
$u(x,t)$ to the bKdV with nonlinearity $f(u)=u^2$ and
\begin{equation*}
b(x,t)=\frac{1}{\epsilon}\lb c\lb
\frac{1}{\sqrt{6}}c_0^\frac{1}{2}h_0\lb x+\frac{c_0}{\epsilon}
t\rb\rb-c_0\rb
\end{equation*}
are related by the transformation
\begin{equation*}
v(y,t)=\frac{4 h_0}{3 c_0}
u\lb\frac{\sqrt{6}}{c_0^\frac{1}{2}h_0}y,
\frac{\sqrt{6}}{c_0^\frac{1}{2}h_0} t\rb.
\end{equation*}
In a wider range of parameters one should add more complicated, in
particular, nonlocal terms to
\eqref{Eqn:KdvGeneralizedWithPotential}. We expect that the
modified equation can still be treated by the methods developed in
this paper.

Similarly, in many other instances in mathematics and the sciences
where the GKdV arises from an approximation of more complicated
systems, the effects of higher order processes can often be
collected into a term of the form $\B(t,x) u$.  Our main result
stated at the end of the next section gives, for long time, an
explicit, leading order description of a solution initially close
to a solitary wave.

We assume that the coefficient $\B$ and nonlinearity $f$ are such
that \eqref{Eqn:KdvGeneralizedWithPotential} has global solutions
for $\Hs{1}$ data and that \eqref{Eqn:KdvGeneralizedWithPotential}
with $\B=0$ possesses solitary wave solutions.  We discuss the
latter assumption in Section
\ref{Section:AssumptionsAndMainResults}.  Here we mention that the
literature regarding well-posedness of the KdV ($\B=0,f(u)=u^2$)
is extensive and well developed. Bona and Smith \cite{BoSm1975}
proved global wellposedness of the KdV in $\Hs{2}$.  See also
\cite{Ka1983}. More recently, Kenig, Ponce, and Vega
\cite{KePoVe1996} have proved local wellposedness in $\Hs{s}$ for
$s\ge -\frac{3}{4}$ and global wellposedness in $\Hs{1}$ for $s\ge
1$. Similar results are available for the gKdV \cite{KePoVe1993}.
More recently, local wellposedness results in negative Sobolev
spaces for the KdV have been extended to global wellposedness
results.  See \cite{CoSt1999,CoKe2001}.  We are not aware of a
wellposedness result for the bKdV in $\Hs{1}$.  Hence, in the next
section we give a global wellposedness result, whose proof (see
Appendix \ref{Appendix:GlobalWellposedness}) uses results of
\cite{KePoVe1993}, and perturbation and energy arguments. We
conjecture that global wellposedness remains true for $b\in C^1$
bounded and subcritical nonlinearities.

Soliton solutions of the KdV equation are known to be orbitally
stable.  Although the linearized analysis of Jeffrey and Kakutani
\cite{JeKa1970} suggested orbital stability, the first nonlinear
stability result was given by Benjamin \cite{Be1972}.  He assumed
smooth solutions and used Lyapunov stability and spectral theory
to prove his results.  Bona \cite{Bo1975} later corrected and
improved Benjamin's result to solutions in $\Hs{2}$. Weinstein
\cite{We1985} used variational methods, avoiding the use of an
explicit spectral representation, and extended the orbital
stability result to the GKdV.  More recently, Grillakis, Shatah,
and Strauss \cite{GrShSt87} extended the Lyapunov method to
abstract Hamiltonian systems with symmetry.  Numerical simulations
of the soliton dynamics for the KdV were performed Bona et al. See
\cite{BoDo86,BoDo91,BoDo95,BoDo96}.

For nonlinear Schr\"{o}dinger and Hartree equations, long-time
dynamics of solitary waves were studied by Bronski and Gerrard
\cite{BrJe2000}, Fr\"{o}hlich, Tsai and Yau \cite{FrTs2003},
Keraani \cite{Ke02}, and Fr\"{o}hlich, Gustafson, Jonsson, and
Sigal \cite{FrGu2003}.  For related results and techniques for the
NLS see also
\cite{BuPe1992,BuSu2003,GaSi2004,RoScSoPreprint,RoScSoPreprintII,TsYa2002I,TsYa2002II,TsYa2002III,SoWe90}.

In our approach we use the fact that the bKdV is a
(non-autonomous, if $b$ depends on time) Hamiltonian system.  As
in the case of the nonlinear Schr\"{o}dinger equation (see
\cite{FrGu2003}), we construct a Hamiltonian reduction of this
original, infinite dimensional dynamical system to a two
dimensional dynamical system on a manifold of soliton
configurations.  The analysis of the general KdV immediately runs
into the problem that the natural symplectic form $\omega$ is not
defined on the tangent space of the soliton manifold.  In the case
of the mKdV ($f(u)=u^3$), the sympletic form is well defined on
the tangent space because of the special structure of the solitary
wave $\Qc$, and hence Dejak and Jonsson \cite{DeJo2004} were able
to prove long time dynamics of solitary waves in this special
case.

To address the problem regarding the symplectic form, we introduce
a family of symplectic forms $\omega_\a$ parametrized by a small
parameter $\a>0$, and approximating $\omega$.  We use the small
parameter to control the errors generated by this approximation.
This approach works, except at one crucial step: the resulting
lower (coercivity) bound on the Hessian of the energy functional
is too weak to close our energy estimates. To remedy this we show
that the weak bound comes from the directions in which we
regularized $\omega$; on the orthogonal complement the lower bound
is sufficiently good. Hence, we decompose a general tangent vector
into "bad" and "good" directions, and use precise information
about the "bad" directions to considerably improve the upper bound
(involving the nonlinearity) and close the energy estimates.

In the next section we formulate our assumptions, state the main
result and describe the organization of the paper.  All $\Lp{2}$
and Sobolev spaces used in this paper, except those in Section
\ref{Section:HessianAndItsProperties}, are real.

\vspace{4mm}\noindent{\bf\large Acknowledgements} \vspace{4mm}\\
We are grateful to J. Bona, R. Pego and Zhou Gang for useful
discussions.

\section{Preliminaries, Assumptions, and the Main Result}
\label{Section:AssumptionsAndMainResults} We begin with the
following global wellposedness result proven in Appendix
\ref{Appendix:GlobalWellposedness}.  See the appendix also for the
definitions of the norms of $\B$ used in the following theorem.
\begin{thm}
\label{Thm:GWP} Let $u_0\in \Hs{1}$.  For small enough
\begin{equation*}
\|\B\|_{XT}:=\LTWXNorm{2}{4}{\infty}{\B}+\LXLTNorm{2}{\infty}{\B}+\|\widehat{\B'}\|_{L_T^2
L_X^1},
\end{equation*}
there is a unique, global solution $u\in C(\R, \Hs{1})$ to
\eqref{Eqn:KdvGeneralizedWithPotential} with $f=u^2$.  With
modification of the norm $\|\B\|_{XT}$, the result continues to
hold for $f=u^3$ and $f=u^4$.
\end{thm}

The bKdV can be written in Hamiltonian form as
\begin{eqnarray}
\dt u=\SympOpInv\HamiltonianWithPotential'(u),
\label{Eqn:KdVVariationalForm}
\end{eqnarray}
where $\HamiltonianWithPotential'$ is the $\Lp{2}$ function
corresponding to the Fr\'{e}chet derivative
$\p\HamiltonianWithPotential$ in the $\Lp{2}$ pairing.  Here the
Hamiltonian $\HamiltonianWithPotential$ is
\begin{eqnarray*}
\HamiltonianWithPotential(u):=\intR{ \frac{1}{2}(\dx
u)^2-F(u)+\frac{1}{2} \B(t,x) u^2},
\end{eqnarray*}
where the function $F$ is the antiderivative of $f$ with $F(0)=0$.
The operator $\dx$ is the anti-self-adjoint operator (symplectic
operator) generating the Poisson bracket
\begin{equation*}
\{F, G\}=\frac{1}{2}\intR{F'(u)\dx G'(u)-G'(u)\dx F'(u)},
\end{equation*}
defined for any $F$, $G$ such that $F', G'\in \Hs{\frac{1}{2}}$.
 The corresponding symplectic form is
\begin{equation*}
\omega(v_1,\, v_2)=\frac{1}{2}\intR{v_1(x)\dx^{-1}
v_2(x)-v_2(x)\dx^{-1} v_1(x)},
\end{equation*}
defined for any $v_1, v_2\in \Lp{1}$.  Here the operator
$\dx^{-1}$ is defined as
\begin{equation*}
\dx^{-1} v(x):=\int_{-\infty}^x v(y)\, dy.
\end{equation*}
Note that $\dx^{-1}\cdot \dx=I$ and, on the space
$\{u\in\Lp{2}\,|\, \intR{u}=0\}$, $\dx^{-1}$ is formally
anti-self-adjoint with inverse $\dx$.  Hence, if
$\intR{v_1(x)}=0$, then $\omega(v_1,\, v_2)=\intR{v_1(x)\dx^{-1}
v_2(x)}$.

Note that if $\B$ depends on time $t$, then equation
\eqref{Eqn:KdVVariationalForm} is non-autonomous.  It is, however,
in the form of a conservation law, and hence the integral of the
solution $u$ is conserved provided $u$ and its derivatives decay
to zero at infinity:
\begin{equation*}
\dd{t}\intR{u}=0.
\end{equation*}
There are also conserved quantities associated to symmetries of
\eqref{Eqn:KdvGeneralizedWithPotential} with $\B=0$. The simplest
such corresponds to time translation invariance and is the
Hamiltonian itself.  This is also true if $\B$ is non-zero but
time independent. If the potential $\B=0$, then
\eqref{Eqn:KdvGeneralizedWithPotential} is also spatially
translation invariant. Noether's theorem then implies that the
flow preserves the momentum
\begin{eqnarray*}
P(u):=\frac{1}{2}\LpNorm{2}{u}^2.
\end{eqnarray*}
In general, when $\B\ne 0$ the temporal and spatial translation
symmetries are broken, and hence, the Hamiltonian and momentum are
no longer conserved. Instead, one has the relations
\begin{align}
\dd{t}\HamiltonianWithPotential(u)&=\frac{1}{2}\intR{(\dt\B) u^2},
\label{Eqn:ConservationHamiltonian}\\
\dd{t}P(u)&=\frac{1}{2}\intR{\B'u^2},
\label{Eqn:ConservationMomentum}
\end{align}
where $\B'(t,x):=\dx \B(t,x)$.  For later use, we also state the
relation
\begin{eqnarray}
\dd{t}\frac{1}{2}\intR{\B u^2}=\intR{\frac{1}{2} u^2\dt\B+\B'\lb u
f(u)-\frac{3}{2}(\dx u)^2-F(u)\rb-\B'' u \dx u}.
\label{Eqn:ConservationPotentialMomentum}
\end{eqnarray}
Assuming (\ref{Eqn:KdvGeneralizedWithPotential}) is well-posed in
$\Hs{2}$, the above equalities are obtained after multiple
integration by parts.  Then, by density of $\Hs{2}$ in $\Hs{1}$,
the equalities continue to hold for solutions in $\Hs{1}$.  To
avoid these technical details, we assume the Hamiltonian flow on
$\Hs{1}$ enjoys (\ref{Eqn:ConservationHamiltonian}),
(\ref{Eqn:ConservationMomentum}) and
(\ref{Eqn:ConservationPotentialMomentum}).

Consider the GKdV, i.e. equation \eqref{Eqn:GKdV}.  Under certain
conditions on $f$, this equation has travelling wave solutions of
the form $\Qc(x-c t)$, where $\Qc$ a positive $\Hs{2}$ function.
Substituting $u=\Qc(x-ct)$ into the GKdV gives the scalar field
equation
\begin{equation}
-\dx^2\Qc+c\Qc-f(\Qc)=0. \label{Eqn:ScalarFieldEquation}
\end{equation}
Existence of solutions to this equation has been studied by
numerous authors. See \cite{St1977, BeLi1983}.  In particular, in
\cite{BeLi1983}, Berestyki and Lions give sufficient and necessary
conditions for a positive and smooth solution $\Qc$ to exist.  We
assume $g:=-c u+f(u)$ satisfies the following conditions:
\begin{enumerate}
\item $g$ is locally Lipschitz and $g(0)=0$, \item
$x^*:=\inf\{x>0\,|\,\int_0^x g(y)\,dy\}$ exists with $x^*>0$ and
$g(x^*)>0$, and \item $\lim_{s\rightarrow 0}\frac{g(s)}{s}\le
-m<0$.
\end{enumerate}
Then, as shown by Berestycki and Lions,
\eqref{Eqn:ScalarFieldEquation} has a unique (modulo translations)
solution $\Qc\in C^2$, which is positive, even (when centred at
the origin), and with $\Qc$, $\dx\Qc$, and $\dx^2\Qc$
exponentially decaying to zero at infinity ($\dx\Qc<0$ for $x>0$).
Furthermore, if $f$ is $C^2$, then the implicit function theorem
implies that $\Qc$ is $C^2$ with respect to the parameter $c$ on
some interval $I_0\subset\Rp$.  We assume that
$x^m\dc^n\Qc\in\Lp{1}$ for $n=1,2,3$, $m=0,1,2$ and that
$\intR{\dc\Qc}\ne 0$. The first assumptions are needed for
continuity and differentiability with respect to $c$  of integrals
containing $\dc^n\Qc$, and the last assumption is made for
convenience.  When $\intR{\vco}=0$, unboundedness of $\dx$ does
not present problems (see \cite{DeJo2004}).

The solitary waves $\Qc$ are orbitally stable if $\delta'(c)>0$,
where $\delta(c)=P(\Qc)$.  See Weinstein \cite{We1985} for
historically the first proof for general nonlinearities. Moreover,
in \cite{GrShSt87}, Grillakis, Shatah and Strauss proved that
$\delta'(c)>0$ is a necessary and sufficient condition for $\Qc$
to be orbitally stable.  In this paper, we assume that $\Qc$ is
stable for all $c$ in some compact interval $I\subset I_0$, or
equivalently that $\delta'(c)>0$ on $I$.  For $f(u)=u^p$, we have
$\delta'(c)=\frac{5-p}{4(p-1)}\LpNorm{2}{Q_{c=1}}^2$, which
implies the well known stability criterion $p<5$ corresponding to
subcritical power nonlinearities.

The scalar field equation for the solitary wave can be viewed as
an Euler-Lagrange equation for the extremals of the Hamiltonian
$\HamiltonianWithoutPotential$ subject to constant momentum
$P(u)$.  Moreover, $\Qc$ is a stable solitary wave if and only if
it is a minimizer of $\HamiltonianWithoutPotential$ subject to
constant momentum $P$. Thus, if $c$ is the Lagrange multiplier
associated to the momentum constraint, then $\Qc$ is an extremal
of
\begin{align}
\ActionAtSoliton(u)&:=\HamiltonianWithoutPotential(u)+c P(u)
\label{DefinitionLS}\\&=\intR{ \frac{1}{2}(\dx u )^2+\frac{1}{2}c
u^2-F(u)},\nonumber
\end{align}
and hence $\ActionAtSoliton'(\Qc)=0$.

The functional $\ActionAtSoliton$ is translationally invariant.
Therefore, $\Qca(x):=\Qc(x-a)$ is also an extremal of
$\ActionAtSoliton$, and $\Qc(x-c t-a)$ is a solitary wave solution
of (\ref{Eqn:KdvGeneralizedWithPotential}) with $\B=0$. All such
solutions form the two dimensional $C^\infty$ manifold of solitary
waves
\begin{equation*}
M_s:=\{\Qca\,|\,c\in I, a\in \R\},
\end{equation*}
with tangent space $\TcaM$ spanned by the vectors
\begin{eqnarray}
\va:=\p_a\Qca=-\dx\Qca\ \mbox{and}\ \vc:=\p_c\Qca,
\label{Eqn:DefinitionOfTangentVectors}
\end{eqnarray}
which we call the translation and normalization vectors.  Notice
that the two tangent vectors are orthogonal.

In addition to the requirements on $\B$ that
(\ref{Eqn:KdvGeneralizedWithPotential}) be globally wellposed, we
assume the potential $\B$ is bounded, twice differentiable, and
small in the sense that
\begin{align}
|\dt^n\dx^m\B|\le\av\tv^n\ev^m, \label{Eqn:AssumptionOnPotential}
\end{align}
for $n=0,1$, $m=0,1,2$, and $n+m\le 2$.  The positive constants
$\av$, $\ev$, and $\tv$ are amplitude, length, and time scales of
the function $\B$. We assume all are less than or equal to one.

Lastly, we make some explicit assumptions on the local
nonlinearity $f$. We require the nonlinearity to be $k$ times
continuously differentiable, with $f^{(k)}$ bounded for some $k\ge
3$ and $f(0)=f'(0)=0$.  These assumptions ensure the Hamiltonian
is finite on the space $\Hs{1}$ and, since $\Qc$ decays
exponentially (see \cite{BeLi1983}), exponential decay of $f(\Qc)$
and $f'(\Qc)$.

We are ready to state our main result.
\begin{thm}
Let the above assumptions hold and assume $\delta'(c)>0$ for all
$c$ in a compact set $I\subset I_0$.  Let $0<s<\frac{1}{2}$. Then,
if $\av\le 1$, $\ev\le 1$, and $\epsilon_0<< (\av\ev)^{2 s}$ are
small enough, there is a positive constant $C$ such that the
solution to (\ref{Eqn:KdvGeneralizedWithPotential}) with an
initial condition $u_0$ satisfying $\inf_{\Qca\in
M_s}\HsNorm{1}{u_0-Q_{c a}}\le\epsilon_0$ can be written as
\begin{eqnarray*}
u(x,t)=Q_{c(t)}(x-a(t))+\e(x,t),
\end{eqnarray*}
where $\HsNorm{1}{\e(t)}\le \O{\av^s\ev^s}$ for all times $t \le
C(\tv+\ev + \av^s\ev^s)^{-1}$.  Moreover, during this time
interval the parameters $a(t)$ and $c(t)$ satisfy the equations
\begin{eqnarray*}
\lb\begin{array}{c}
  \dot{a} \\
  \dot{c}
\end{array}\rb&=&
\lb\begin{array}{c}
c-\B(t,a)\\
0
\end{array}\rb+\O{\av^{2 s}\ev^{2 s}+\av\ev^2},
\end{eqnarray*}
where $c$ is assumed to lie in the compact set $I$.
\label{MainThm}
\end{thm}
\begin{proof}[Sketch of Proof and Paper Organization]
To realize the Hamiltonian reduction we decompose functions in a
neighbourhood of the soliton manifold $M_s$ as
\begin{equation*}
u=\Qca+\e
\end{equation*}
with $\e$ symplectically orthogonal to $\TcaM$, i.e.
$\e\bot\dx^{-1}\TcaM$.  Unfortunately, since
$\dx^{-1}\TcaM\not\subseteq\Lp{2}$, such a decomposition is
ill-defined for $\e\in\Hs{1}$.  To overcome this difficulty we
construct in Section \ref{Section:Modulation} an approximate
symplectic form
\begin{equation*}
\omega(v_1,\, v_2)=\frac{1}{2}\intR{v_1(x)\lb \AnisoRegSympOp
-\AnisoRegSympOp^*\rb v_2(x)},
\end{equation*}
where $\a>0$ and $\AnisoRegSympOp$ is a bounded operator
regularizing the unbounded operator $\dx^{-1}$ in certain
directions.  We show that there is an $\epsilon_0>0$ such that if
the solution $u$ satisifes the estimate
$\inf_{\Qca}\HsNorm{1}{u-\Qca}<\epsilon_0$, then there are unique
$C^1$ functions $a(u)$ and $c(u)$ such that $u=Q_{c(u) a(u)}+\e$
with $\e\in\lb\AnisoRegSympOp\TcaM\rb^\bot$.

With the knowledge that the symplectic decomposition exists, we
substitute $u=\Qca+\e$ into the bKdV
\eqref{Eqn:KdvGeneralizedWithPotential} and split the resulting
equation according to the decomposition
\begin{equation*}
\Lp{2}=\AnisoRegSympOp\TcaM\oplus\lb\AnisoRegSympOp\TcaM\rb^{\bot}
\end{equation*}
to obtain equations for the parameters $c$ and $a$, and an
equation for the (infinite dimensional) fluctuation $\e$.  In
Sections \ref{Section:HessianAndItsProperties} and
\ref{Section:Positivity}, we establish spectral properties and an
anisotropic lower bound of the Hessian $\ActionAtSoliton''$ on the
space $\lb\AnisoRegSympOp\TcaM\rb^{\bot}$. Using these properties
we orthogonally decompose $\e$ into a "bad" $\emin$ part and a
"good" part $\ep$, where $\emin$ is colinear with the minimizer
$\eta$ of $\LpNorm{2}{\e}^{-2}\ip{\e}{\Lca\e}$. In Section
\ref{Section:Projection} we isolate the leading order terms in the
equations for $a$ and $c$ and estimate the remainder, including
all terms containing $\e=\emin+\ep$.  We use the special
properties of the minimizer $\eta$ to obtain a better estimate on
the nonlinear terms containing $\emin$.

The proof that $\HsNorm{1}{\e}$ is sufficiently small is the final
ingredient in the proof of the main theorem.  The remaining
sections concentrate on proving this crucial result.  We employ a
Lyapunov method and in Section \ref{Section:LyapDeriv} we
construct the Lyapunov function $M_c$ and prove an estimate on its
time derivative. This estimate is later time maximized over an
interval $[0,T]$, and integrated to obtain an upper bound on $M_c$
involving the time $T$ and the norms of $\emin$ and $\ep$. This
anisotropic upper bound is considerably better than an isotropic
bound.  We combine this upper bound with the anisotropic lower
bound $M_c\gtrsim C\a\HsNorm{1}{\emin}+C\HsNorm{1}{\ep}$, which
follows from the results of Section \ref{Section:Positivity}, and
obtain and inequality involving the norms of $\emin$ and $\ep$.
Here $\a$ is the small regularization parameter mentioned in the
introduction and will be taken small, in fact $\a=(\av\ev)^s$.
This inequality implies upper bounds on $\HsNorm{1}{\emin}$ and
$\HsNorm{1}{\ep}$, provided $\HsNorm{1}{\e(0)}$ is small enough,
via the standard argument given Section
\ref{Section:BoundOnFluct}.  We substitute this bound into the
bound appearing in the dynamical equation for $a$ and $c$, and
take $\av\ev$ and $\epsilon_0$ small enough so that all
intermediate results hold to complete the proof.
\end{proof}

\section{Modulation of Solutions}
\label{Section:Modulation} As stated in the previous section, we
begin the proof by decomposing the solution of
\eqref{Eqn:KdvGeneralizedWithPotential} into a modulated solitary
wave and a fluctuation $\e$:
\begin{eqnarray}
u(x,t)=Q_{c(t)a(t)}(x)+\e(x,t),
\label{EquationWithUErrorQDecomposition}
\end{eqnarray}
with $a$, $c$, and $\e$ fixed by an orthogonality condition, which
we now describe.  Ideally we would like to take $\e$ orthogonal to
$\SympOp\TcaM$, where $\SympOp$ is the symplectic operator defined
on absolutely continuous functions $g$ as
\begin{equation*}
\SympOp:g\mapsto\int_{-\infty}^x g(y)\,dy.
\end{equation*}
It is easy to see that $\SympOpInv\SympOp=I$ and if
$\lim_{x\rightarrow-\infty} g(x)=0$, then $\SympOp\SympOpInv=I$.
The problem here is that $\SympOp\TcaM\nsubseteq\Lp{2}$. More
precisely, while $\SympOp\va=-\Qca\in\Lp{2}$ we have that in
general $\SympOp\vc\not\in\Lp{2}$.
 In fact, if $f(u)=u^p$, then
\begin{equation*}
\SympOp\vc|_{a=0}=\frac{x\Qc}{2 c}+\frac{3-p}{2
c(p-1)}\int_{-\infty}^y\Qc(y)\, dy
\end{equation*}
and therefore
\begin{equation*}
\lim_{x\rightarrow\infty}\SympOp\vc=\frac{3-p}{2
c(p-1)}\int_{-\infty}^\infty\Qc(y)\,dy.
\end{equation*}
Since $\Qc$ is positive, $\SympOp\vc$ is not an $\Lp{2}$ function
if $p\ne 3$.  We remark that if $p=3$, then there are no problems.
This case is in the special class of nonlinearities considered in
\cite{DeJo2004}.

Our remedy to the above problem is to "regularize" the symplectic
operator $\SympOp$.  Let $\NullProj$ be the $\Lp{2}$ orthogonal
projection onto the subspace spanned by the translation vector
$\va$, and let $\NullProjBar$ be its orthogonal complement. Then
we define the anisotropic regularization $\AnisoRegSympOp$ of
$\SympOp$ as
\begin{equation*}
\AnisoRegSympOp:=\SympOp\NullProj+\RegSympOp\NullProjBar,
\end{equation*}
where $\RegSympOp:=(\dx+\a)^{-1}$ is the regularization of
$\SympOp$. We do not regularize in the direction of $\va$ since
$\SympOp$ is well behaved on this vector.

For $\RegSympOp$ to exist, the parameter $\a$ must lie in the
resolvent set $\rho(\dx)=\C\setminus i\R$, and in such a case
$\RegSympOp$ acts explicitly as
\begin{equation*}
\RegSympOp:g\mapsto\int_{-\infty}^x g(y)e^{\a(y-x)}\, dy
\end{equation*}
on all $\Lp{2}$ functions $g$.  The lemma below, proven in
Appendix \ref{Appendix:PropertiesOfRegSympOpAndRestriction},
collects some properties of $\RegSympOp$, which will be used in
the course of proving the main result.
\begin{lemma}
\label{Lemma:RegularizedSymplecticOperatorProperties} Let
$\phi,\psi\in\Lp{1}\cap\Lp{2}$ and $\a\in\Rp$.  Then we have
\begin{enumerate}
\item  The operator $\RegSympOp$ commutes with $\dx$ and spatial
translation ${\cal S}_a:f(x)\rightarrow f(x-a)$; that is,
$\dx\RegSympOp=\RegSympOp\dx=I-\a\RegSympOp$ and ${\cal
S}_a\RegSympOp=\RegSympOp{\cal S}_a$.
    \label{LemmaItem:RegularizedSymplecticOperatorCommutativity}
\item $\LpNorm{\infty}{\RegSympOp \phi}\le\LpNorm{1}{\phi}$.
    \label{LemmaItem:RegularizedSymplecticOperatorSupNorm}
\item There is a constant $C$ such that
$\LpNorm{2}{\RegSympOp\phi}\le
C\a^{-\frac{1}{2}}\LpNorm{1}{\phi}$.
    \label{LemmaItem:RegularizedSymplecticOperatorLIINorm}
\item If $x\phi\in\Lp{1}$, then
    $\LpNorm{2}{x\RegSympOp\phi}\le C\lb\a^{-\frac{3}{2}}\LpNorm{1}{\phi}
    +\a^{-\frac{1}{2}}\LpNorm{1}{x\phi}\rb$.
    \label{LemmaItem:RegularizedSymplecticOperatorXPhiLIINorm}
\item If $x\phi,x\psi\in\Lp{1}$, then
    $\left|\ip{\phi}{\RegSympOp\psi}-\ip{\phi}{\SympOp\psi}\right|\le\a\lb\LpNorm{1}{\phi}\LpNorm{1}{x\psi}+\LpNorm{1}{x\phi}\LpNorm{1}{\psi}\rb$ and, in
    particular,
    \begin{equation*}
    \left|\ip{\phi}{\RegSympOp\phi}-\frac{1}{2}\lb\intR{\phi}\rb^2\right|\le
    2\a\LpNorm{1}{\phi}\LpNorm{1}{x\phi}.
    \end{equation*}
    \label{LemmaItem:RegularizedSymplecticOperatorInner}
\item If
    $x\phi, x^2\phi\in\Lp{1}$, then
    \begin{equation*}
    \LpNorm{2}{\RegSympOp\phi}=\frac{\pi}{\a}\lb\intR{\phi}\rb^2+\O{1}.
    \end{equation*}
    \label{LemmaItem:RegularizedSymplecticOperatorLIILeading}
\end{enumerate}
\end{lemma}

For $\a$ small, the above lemma implies that the properties of
$\SympOp$ and $\RegSympOp$ are similar.  Thus, we require in
\eqref{EquationWithUErrorQDecomposition} that
\begin{equation}
\e\bot\AnisoRegSympOp\TcaM. \label{Cond:Orthogonality}
\end{equation}
The existence and uniqueness of parameters $a$ and $c$ such that
$\e=u-\Qca$ satisfies \eqref{Cond:Orthogonality} follows from the
next lemma concerning a restriction of $\AnisoRegSympOp$ and the
implicit function theorem.

The restriction $\AnisoRest$ of $\AnisoRegSympOp$ to the tangent
space $\TcaM$ is defined by the equation $\AnisoRest
\TMProj=\TMProj\AnisoRegSympOp \TMProj$, where $\TMProj$ is the
orthogonal projection onto $\TcaM$.  In the natural basis
$\{\va,\vc\}$ of the tangent space $\TcaM$, the matrix
representation of $\AnisoRest$ is $N^{-1}\AnisoRestNormlzd$, where
\begin{eqnarray*}
 N&:=&\lb\begin{array}{cc}
  \LpNorm{2}{\va}^2 & 0 \\
  0 & \LpNorm{2}{\vc}^2
\end{array}\rb
\end{eqnarray*}
and
\begin{eqnarray}
 \AnisoRestNormlzd&:=&
\lb\begin{array}{cc}
  \ip{\va}{\AnisoRegSympOp\va} & \ip{\vc}{\AnisoRegSympOp\va} \\
  \ip{\va}{\AnisoRegSympOp\vc} & \ip{\vc}{\AnisoRegSympOp\vc}
\end{array}\rb.
\label{Eqn:DefinitionOfOiN}
\end{eqnarray}
Notice that the matrix $\AnisoRestNormlzd$ depends on the base
point $\Qca$, and hence on the parameters $a$ and $c$.
\begin{lemma}
\label{Lemma:InvertibilityOfOiOmegaBound} If $\delta'(c)> 0$ on a
compact set $I\subset\Rp$ and $\a<<\InfI{\delta'}:=\inf_I
\delta'$, then $\AnisoRestNormlzd$ is invertible for all $c\in I$
and $a\in\R$, and
\begin{eqnarray} \AnisoRestNormlzdInv=\frac{1}{\delta'(c)^2}\lb\begin{array}{cc}
\frac{1}{2}\lb\intR{\vc}\rb^2 & \delta'(c)\\
-\delta'(c) & 0\\
\end{array}\rb+\O{\frac{\a}{\InfI{\delta'}}}.
\label{Eqn:LeadingOrderExpressionOfSymplecticInverse}
\end{eqnarray}
Hence, $\|\AnisoRestNormlzdInv\|=\O{\InfI{\delta'}^{-2}}$ for
$\InfI{\delta'}$ small.
\end{lemma}
\begin{proof}
We use the relations $\va=-\dx\Qca$,
$\AnisoRegSympOp\va=\SympOp\va$,
$\AnisoRegSympOp\vc=\RegSympOp\vc$, anti-self-adjointness of
$\dx$, and $\SympOp\dx=I$ to simplify the matrix
$\AnisoRestNormlzd$ into
\begin{equation*}
\AnisoRestNormlzd=\lb
\begin{array}{cc}
  0 & -\ip{\vc}{\Qca}\\
  \ip{\Qca}{\dx\RegSympOp\vc} & \ip{\vc}{\RegSympOp\vc}
\end{array}
\rb.
\end{equation*}
Next, using statements
\ref{LemmaItem:RegularizedSymplecticOperatorCommutativity} and
\ref{LemmaItem:RegularizedSymplecticOperatorInner} of the previous
Lemma, we separate the leading order part of $\AnisoRestNormlzd$
from the higher order parts, and use the relation
$\delta'(c)=\ip{\Qca}{\vc}$ to obtain that
\begin{eqnarray*}
\AnisoRestNormlzd=\lb
\begin{array}{cc}
0 & -\delta'(c)\\
\delta'(c) & \frac{1}{2}\lb\intR{\vc}\rb^2
\end{array}
\rb+ \lb
\begin{array}{cc}
0 & 0\\
\a\ip{\Qca}{\RegSympOp\vc} & R
\end{array}
\rb,
\end{eqnarray*}
where $|R|\le 2\a\sup_I\LpNorm{1}{\vc} \LpNorm{1}{x\vc}$. With
$\Qca$ and $\vc$ exponentially decaying, the estimate
$\left|\ip{\Qca}{\RegSympOp\vc}\right|\le
\LpNorm{1}{\Qca}\LpNorm{1}{\vc}$ is clear from the properties of
$\RegSympOp$.  Thus, if
$\a\le\frac{1}{2}\InfI{\delta'}\lb\sup_I\LpNorm{1}{\Qca}\LpNorm{1}{\vc}\rb^{-1}$,
then the determinant
\begin{eqnarray*}
\det
\AnisoRestNormlzd=\delta'(c)^2+\a\delta'(c)\ip{\Qca}{\RegSympOp\vc}\ge\frac{1}{2}\InfI{\delta'}^2,
\end{eqnarray*}
and hence it is nonzero for all $c\in I$ and $a\in\R$.  The
coadjoint formula and the above estimate give
\eqref{Eqn:LeadingOrderExpressionOfSymplecticInverse}. The
estimate of $\|\AnisoRestNormlzdInv\|$ follows from
\eqref{Eqn:LeadingOrderExpressionOfSymplecticInverse} and the
assumption that $\intR{\vc}\ne 0$.
\end{proof}

Given $\varepsilon>0$, define the tubular neighbourhood
$U_{\varepsilon}:=\{u\in\Lp{2}\,|\,\inf_{(c,\,a)\in
I\times\R}\LpNorm{2}{u-\Qca}<\varepsilon\}$ of the solitary wave
manifold $M_s$ in $\Lp{2}$.
\begin{prop}
\label{Prop:ExistenceOfDecomposition} Let $I\subset\Rp$ be a
compact interval such that $c\mapsto\Qca$ is $C^1(I)$.  Then, if
$\a<<\InfI{\delta'}$, there exists a positive number
$\varepsilon=\varepsilon(I)=\O{\a^\frac{1}{2}\InfI{\delta'}^{4}}$
and unique $C^1$ functions $ a:U_{\varepsilon}\rightarrow\Rp$ and
$c:U_{\varepsilon}\rightarrow I$, dependent on $\a$ and $I$, such
that
\begin{equation*}
\ip{Q_{c(u)a(u)}-u}{\AnisoRegSympOp\zeta^{tr}_{c(u)a(u)}}=0\
\mbox{and}\
\ip{Q_{c(u)a(u)}-u}{\AnisoRegSympOp\zeta_{c(u)a(u)}^n}=0
\end{equation*}
for all $u\in U_{\varepsilon}$.  Moreover, there is a positive
real number $C=C(I)$ such that
\begin{equation}
\HsNorm{1}{u-Q_{c(u) a(u)}}\le C\a^{-\frac{1}{2}}\inf_{\Qca\in
M_s}\HsNorm{1}{u-\Qca} \label{Ineq:InitialConditionIFT}
\end{equation}
for all $u\in U_{\varepsilon}\cap\Hs{1}$.
\end{prop}
\begin{proof}
Let $\mu:=\lb a\ c\rb^T$ and define $F:\Lp{2}\times\Rp\times
I\rightarrow\R^2$ by
\begin{eqnarray*}
F:(u,\mu)\mapsto\lb\begin{array}{c}
  \ip{\Qca-u}{\AnisoRegSympOp\va}\\
  \ip{\Qca-u}{\AnisoRegSympOp\vc}
\end{array}\rb.
\end{eqnarray*}
The proposition is equivalent to solving $F(u,g(u))=0$ for a $C^1$
function $g$.  Observe that $F$ is $C^1$ and $F(\Qca,\mu)$=0.  To
apply the implicit function theorem it suffices to check that
$\p_\mu F(\Qca,\mu)$ is invertible.  Then the there exists an open
ball $B_{\varepsilon}(\Qca)$ of radius $\varepsilon$ with centre
$\Qca$, and a unique function
$g_{\a\Qca}:B_\varepsilon(\Qca)\rightarrow\Rp\times I$, such that
$F(u,g_{\a\Qca}(u))=0$ for all $u\in B_{\varepsilon}(\Qca)$. Since
$\p_\mu F(\Qca,\mu)=\AnisoRestNormlzd$, the invertibility of
$\p_\mu F(\Qca,\mu)$ follows from Lemma
\ref{Lemma:InvertibilityOfOiOmegaBound} provided $\a$ is small
enough.  The radius of the balls $B_\varepsilon(\Qca)$ depend on
the parameters $c$, $a$ and $\a$. To obtain an estimate of the
radius, and to show that we can take $\varepsilon$ independent of
the parameters $c$ and $a$, we give a proof of the existence of
the functions $g_{\a\Qca}$ using the contraction mapping principle
(just as in the proof of the implicit function theorem).

Expand $F(u,\mu)$ to linear order in $\mu$ around $\mu_0=(a\,
c)^T$:
\begin{equation}
F(u,\mu)=F(u,\mu_0)+\p_\mu F(u,\mu_0)(\mu-\mu_0)+R(u,\mu)
\label{Eqn:ContractionOne},
\end{equation}
where $R(u,\mu)=\frac{1}{2}\p_\mu^2
F(u,(1-\lambda)\mu_0+\lambda\mu)(\mu-\mu_0)^2$ for some
$\lambda\in [0,1]$.  The operator $\p_\mu F(u,\mu_0)$ is computed
to be $\p_\mu F(u,\mu_0)=\AnisoRestNormlzd+A$, where
\begin{equation*}
A:=\lb
\begin{array}{cc}
\ip{\Qca-u}{\p_a\Qca} & \ip{\Qca-u}{\dc\Qca}\\
\ip{\Qca-u}{\p_a\RegSympOp\vc} & \ip{\Qca-u}{\p_c\RegSympOp\vc}.
\end{array}
\rb
\end{equation*}
If $u\in B_\varepsilon(\Qca)$, where $\varepsilon$ remains to be
chosen, then the properties of $\RegSympOp$ imply that $\|A\|\le
C\a^{-\frac{1}{2}}\varepsilon$.  Thus, since $\AnisoRestNormlzd$
is invertible, if $\varepsilon << (C\sup_I
\|\AnisoRestNormlzdInv\|)^{-1}\a^\frac{1}{2}$, then $\p_\mu
F(u,\mu_0)$ is invertible and $\|[\p_\mu F(u,\mu_0)]^{-1}\|\le
C\sup_I\|\AnisoRestNormlzdInv\|$. Hence, given $u$, $F(u,\mu)=0$
has a solution $\mu$ if and only if
\begin{equation*}
\mu=H(\mu):=\mu_0-[\p_\mu F(u,\mu_0)]^{-1}\lb F(u,\mu_0)+R(u,\mu)
\rb
\end{equation*}
has a solution $\mu$.  The latter is equivalent to the function
$H$ having a fixed point.  This is guaranteed by the contraction
mapping principle if $H$ is a strict contraction from some ball
$B_\rho(\mu_0)$ to $B_\rho(\mu_0)$.

Say $\mu\in B_\rho(\mu_0)$, where $\rho$ remains to be chosen, and
consider the bound
\begin{equation*}
\|H(\mu)-\mu_0\|\le
C\sup_I\|\AnisoRestNormlzdInv\|\|F(u,\mu_0)+R(u,\mu)\|.
\end{equation*}
After subtracting $F(\Qca,\mu_0)=0$ from $F(u,\mu_0)$ and using
the mean value theorem, the above becomes
\begin{eqnarray*}
\|H(\mu)-\mu_0\|\le C\sup_I\|\AnisoRestNormlzdInv\|\lb \|\p_u
F((1-\lambda_1)\Qca+\lambda_1
u,\mu_0)\|\,\|u-\Qca\|\right.\\+\frac{1}{2}\left.\| \p_\mu^2
F(u,(1-\lambda_2)\mu_0+\lambda_2\mu) \|\,\|\mu-\mu_0\|^2 \rb
\end{eqnarray*}
for some $\lambda_1$, $\lambda_2\in [0,1]$.  Again, using the
properties of $\RegSympOp$ we find that
\begin{align}
\|\p_u F((1-\lambda_1)\Qca+\lambda_1 u, \mu_0)\|&\le
C\a^{-\frac{1}{2}}\nonumber\ \\ \|\p_\mu^2
F(u,(1-\lambda_2)\mu_0+\lambda_2\mu\|&\le
C(1+\a^{-\frac{1}{2}})\label{Eqn:ContractionTwo}
\end{align}
for all $\mu_0,\mu\in \Rp\times I$ and $u\in B_\varepsilon(\Qca)$.
Thus, if $\rho<1$, then $\|H(\mu)-\mu_0\|\le
C\sup_I\|\AnisoRestNormlzdInv\|\lb
\a^{-\frac{1}{2}}\varepsilon+\rho^2 \rb$.  Taking $\varepsilon <<
(C\sup_I \|\AnisoRestNormlzdInv\|)^{-1}\a^\frac{1}{2}\rho$ and
$\rho << (C\sup_I\|\AnisoRestNormlzdInv\|)^{-1}$ implies $H$ maps
$B_\rho(\mu_0)$ into $B_\rho(\mu_0)$.

Let $\mu_1,\mu_2\in B_\rho(\mu_0)$ and consider the bound
\begin{equation*}
\|H(\mu_2)-H(\mu_1)\|\le
C\sup_I\|\AnisoRestNormlzdInv\|\,\|R(u,\mu_2)-R(u,\mu_1)\|
\end{equation*}
or, using the mean value theorem, the derivative of
\eqref{Eqn:ContractionOne} with respect to $\mu$, and the mean
value theorem again,
\begin{eqnarray*}
\|H(\mu_2)-H(\mu_1)\|\le
C\sup_I\|\AnisoRestNormlzdInv\|\,\|\p_\mu^2
F(u,(1-\lambda_2)[(1-\lambda_1)\mu_1+\lambda_1\mu_2]+\lambda_2\mu_0)\|\\\times
\|(1-\lambda_1)\mu_1+\lambda_1\mu_2-\mu_0\|\,\|\mu_2-\mu_1\|
\end{eqnarray*}
for some $\lambda_1,\lambda_2\in[0,1]$.  Using
\eqref{Eqn:ContractionTwo} and
$\|(1-\lambda_1)\mu_1+\lambda_1\mu_2-\mu_0\|<\rho$ then gives
\begin{equation*}
\|H(\mu_2)-H(\mu_1)\|\le
C\sup_I\|\AnisoRestNormlzdInv\|(1+\a^{-\frac{1}{2}}\varepsilon)\rho\|\mu_2-\mu_1\|.
\end{equation*}
Thus, with the above choices of $\varepsilon$ and $\rho$, $H$ is a
strict contraction.  We conclude that the radii of the balls
$B_\varepsilon(\Qca)$ can be taken independent of $a$, $c$ (but
dependent on $I$) and
$\varepsilon=\O{\a^\frac{1}{2}\InfI{\delta'(c)}^{4}}$.

The above argument shows that there exists balls $\{
B_{\varepsilon}(\Qca)\, |\, a\in\Rp, c\in I\}$ with radius
$\varepsilon$ dependent only on the parameter $\a$ and the compact
set $I$. Notice that $U_{\varepsilon}=\bigcup\{
B_{\varepsilon}(\Qca)\, |\, a\in\Rp, c\in I\}$.  Pasting the $C^1$
functions $g_{\a\Qca}$ together into a $C^1$ function $g_{\a
I}:U_{\varepsilon}\rightarrow \Rp\times I$ gives the required
$C^1$ functions $a(u)$ and $c(u)$. Uniqueness follows from the
uniqueness of each of the functions $g_{\a\Qca}$.

Let $u\in U_{\varepsilon}$, $c\in I$ and $a\in \R$, and consider
the equation
\begin{equation*}
u-Q_{c(u)a(u)}=u-\Qca+\Qca-Q_{c(u)a(u)}.
\end{equation*}
Clearly, inequality \eqref{Ineq:InitialConditionIFT} will follow
if $\HsNorm{1}{\Qca-Q_{c(u)a(u)}}\le C\HsNorm{1}{u-\Qca}$ for some
positive constant $C$.  Since the derivatives $\p_c\Qca$ and
$\p_a\Qca$ are uniformly bounded in $\Hs{1}$ over $I\times\R$, the
mean value theorem gives that $\HsNorm{1}{\Qca-Q_{c(u)a(u)}}\le
C\|(a\, c)^T-(a(u)\,c(u))^T\|$, where the constant $C$ does not
depend on $c$, $a$, or $\a$.  The relations $g_{\a I}(\Qca)=(a\,
c)^T$ and $g_{\a I}(u)=(a(u)\, c(u))^T$ then imply
$\HsNorm{1}{\Qca-Q_{c(u)a(u)}}\le C\|g_{\a I}(\Qca)-g_{\a
I}(u)\|$.  Again, we appeal to the mean value theorem and obtain
\eqref{Ineq:InitialConditionIFT}, using the properties of
$\RegSympOp$ and that $\p_u g_{\a I}=-\p_\mu F^{-1}\p_u F$ is
uniformly bounded in the parameters $c\in I$, $a\in \Rp$ and $u\in
U_\varepsilon$.
\end{proof}


\section{Spectral Properties of the Hessian $\p^2\ActionAtSoliton$}
\label{Section:HessianAndItsProperties} The Hessian
$\p^2\ActionAtSoliton$ at $\Qca$ in the $\Lp{2}$ pairing is
computed to be the unbounded operator
\begin{align}
\Lca&:=-\dx^2+c-f'(\Qca), \label{Eqn:Hessian}
\end{align}
defined on $\Lp{2}$ with domain $\Hs{2}$.  We extend this operator
to the corresponding complex spaces.
\begin{prop}  \label{Prop:Spectrum}
The self-adjoint operator $\Lca$ has the following properties.
{\begin{enumerate}
    \item $\Lca\va=0$ and $\Lca\vc=-\Qca$.
    \item All eigenvalues of $\Lca$ are simple, and
    $\Null{\Lca}=\Span{\va}$.
    \item $\Lca$ has exactly one negative eigenvalue.
    \item The essential spectrum is $[c,\infty)\subset\Rp$.
    \item $\Lca$ has a finite number of eigenvalues in $(-\infty,
    c)$.
\end{enumerate}}
\end{prop}
\begin{proof}
Recall that the vectors $\va:=-\dx\Qca$ and $\vc:=\dc\Qca$ are in
the Sobolev space $\Hs{2}$.  Thus, relations $\Lca\va=0$ and
$\Lca\vc=-\Qca$ make sense, and are obtained by differentiating
$\ActionAtSoliton'(\Qca)=0$ with respect to $a$ and $c$.  The
first relation above proves that $\va$ is a null vector.

Say $\zeta,\eta\in \Hs{2}$ are linearly independent eigenvectors
of $\Lca$ with the same eigenvalue. Then, since $\Lca$ is a second
order linear differential operator without a first order
derivative, the Wronskian
\begin{eqnarray*}
W(\eta,\zeta)=\zeta\dx \eta-\eta\dx\zeta
\end{eqnarray*}
is a non-zero constant.  With $\eta$ and $\zeta$ both in $\Hs{2}$
however, the limit $\lim_{x\rightarrow \infty} W(\eta,\zeta)$ is
zero. This contradicts the non vanishing of the Wronskian, and
hence all eigenvalues of $\Lca$ are simple and, in particular,
$\Null{\Lca}=\Span{\va}$.

Next we prove that the operator $\Lca$ has exactly one negative
eigenvalue using Sturm-Liouville theory on an infinite interval.
Recall that the solitary wave $\Qca(x)$ is a differentiable
function, symmetric about $x=a$ and monotonically decreasing if
$x>a$. This implies that the null vector $\va$, or equivalently,
the derivative of $\Qca$ with respect to $x$, has exactly one root
at $x=a$. Therefore, by Sturm-Liouville theory, zero is the second
eigenvalue and there is exactly one negative eigenvalue.

We use standard methods to compute the essential spectrum.  Since
the function $f'(\Qca(x))$ is continuous and decays to zero at
infinity, the bottom of the essential spectrum begins at
$\lim_{x\rightarrow \infty} (c-f'(\Qca(x)))=c$ and extends to
infinity: $\sigma_{ess}(\Lca)=[c,\infty)$. Furthermore, the bottom
of the essential spectrum is not an accumulation point of the
discrete spectrum since $f'(\Qca(x))$ decays faster than $x^{-2}$
at infinity.  Hence, there is at most a finite number of
eigenvalues in the interval $(-\infty,c)$.  For details see
\cite{ReSiI, ReSiIV, GuSi2003}.
\end{proof}

\section{Anisotropic Coercivity of the Hessian $\Lca$ on $\lb\AnisoRegSympOp\TcaM\rb^\bot$}
\label{Section:Positivity} In this section we prove strict
positivity of the Hessian $\Lca$ on the orthogonal complement of
the 2-dimensional space
$\AnisoRegSympOp\TcaM=\Span{\Qca,\,\RegSympOp\vc}$.  This result
is a crucial ingredient in the proof of the bound on the
fluctuation $\e$.
\begin{prop}
\label{Prop:PositivityHessian} Assume $\delta'(c)>0$ on the
compact set $I\subset I_0$.  The following statements hold if
$\a>0$ is small enough and $\e\bot\AnisoRegSympOp\TcaM$.
\begin{enumerate}
    \item There are positive real numbers $C_1$ and $C_2$, independent of $\a$, and a function $\varrho(\a)$ satisfying $C_1\a\le \varrho(\a)\le
    C_2\a$ such that $\ip{\Lca\e}{\e}\ge\varrho(\a)\HsNorm{1}{\e}^2$ for all $c\in I$ and $a\in \R$.
    \label{Item:FirstCoer}
    \item The infimum $\inf\{\ip{\Lca\e}{\e}\, |\,
    \e\bot\AnisoRegSympOp\TcaM\ \mbox{and}\ \LpNorm{2}{\e}=1
\}$ is attained and the unique minimizer $\eta$ is of the form
$\eta=\gamma\va+\etabot$, where
$\HsNorm{1}{\etabot}=\O{\a^\frac{1}{2}}$ and $\gamma=\O{1}$.
    \item Let $\eta$ be as above and let $\ep:=\e-\ip{\e}{\eta}{\eta}$.
    There exists a positive real number $C_3$ independent of $\a$,
    such that
    $\ip{\Lca\ep}{\ep}\ge C_3\HsNorm{1}{\ep}^2$ (notice that $\a$ enters $\ep$ through the minimizer
    $\eta$).
\item The Hessian $\Lca$ is anisotropically coercive on
$\lb\AnisoRegSympOp\TcaM\rb^\bot$; that is,
\begin{equation*}
\ip{\Lca\e}{\e}\ge C_3\HsNorm{1}{\ep}^2+C\a\HsNorm{1}{\emin}^2.
\end{equation*}
\end{enumerate}
\end{prop}
\begin{proof}
Define the set $X:=\{ \e\in \Hs{1}\,
|\,\e\bot\AnisoRegSympOp\TcaM\ \mbox{and}\ \LpNorm{2}{\e}=1 \}$.
Our first step is to prove an upper bound on
$\inf_X\ip{\Lca\e}{\e}$.  We do this by computing
$\ip{\Lca\e}{\e}$ for the test function
\begin{eqnarray*}
\e:=\lambda_1\va+\lambda_2\RegSympOp\vc+\lambda_3\Qca,
\end{eqnarray*}
where $\lambda_1$, $\lambda_2$, and $\lambda_3$ are chosen to
satisfy $\ip{\e}{\Qca}=0$, $\ip{\e}{\RegSympOp\vc}=0$, and
$\LpNorm{2}{\e}^2=1$.  These conditions imply $\e\in X$, and,
after substituting $\e$ with its definition, have the form
\begin{align}
\lambda_2\ip{\Qca}{\RegSympOp\vc}+\lambda_3\LpNorm{2}{\Qca}^2&=0,
\label{Eqn:TestFunctionFirstCondition}\\
\lambda_1\ip{\va}{\RegSympOp\vc}+\lambda_2\LpNorm{2}{\RegSympOp\vc}^2+\lambda_3\ip{\Qca}{\RegSympOp\vc}&=0,
\label{Eqn:TestFunctionSecondCondition}
\end{align}
and
\begin{align}
\lambda_1^2\LpNorm{2}{\va}^2&+\lambda_2^2\LpNorm{2}{\RegSympOp\vc}^2+\lambda_3^2\LpNorm{2}{\Qca}^2\nonumber\\&+2\lambda_1\lambda_2\ip{\va}{\RegSympOp\vc}+2\lambda_2\lambda_3\ip{\Qca}{\RegSympOp\vc}=1.
\label{Eqn:TestFunctionThirdCondition}
\end{align}
Equation \eqref{Eqn:TestFunctionSecondCondition} can be solved for
$\lambda_1$ when $\ip{\va}{\RegSympOp\vc}$ is not zero.  A
straightforward computation using antisymmetry of $\dx$ and
statements
\ref{LemmaItem:RegularizedSymplecticOperatorCommutativity} and
\ref{LemmaItem:RegularizedSymplecticOperatorInner} of Lemma
\ref{Lemma:RegularizedSymplecticOperatorProperties} gives that
\begin{eqnarray}
\ip{\va}{\RegSympOp\vc}=\delta'(c)+\O{\a}.
\label{Est:VaRegSympOpVc}
\end{eqnarray}
Thus, if $\a$ is small enough, then $\ip{\va}{\RegSympOp\vc}$ is
non-zero.  We substitute for $\lambda_1$ in
\eqref{Eqn:TestFunctionThirdCondition} using
\eqref{Eqn:TestFunctionSecondCondition} and then use
\eqref{Eqn:TestFunctionFirstCondition} to substitute for
$\lambda_3$ to obtain
\begin{eqnarray}
\left[\frac{\LpNorm{2}{\va}^2}{\ip{\va}{\RegSympOp\vc}^2}\lb\LpNorm{2}{\RegSympOp\vc}^2-\frac{\ip{\Qca}{\RegSympOp\vc}^2}{\LpNorm{2}{\Qca}^2}\rb^2-\LpNorm{2}{\RegSympOp\vc}^2+\frac{\ip{\RegSympOp\vc}{\Qca}^2}{\LpNorm{2}{\Qc}^2}\right]\lambda_2^2=1.
\label{Eqn:Lambda2}
\end{eqnarray}
This equation, the relations
$\LpNorm{2}{\RegSympOp\vc}^2=\frac{\pi}{\a}\lb\intR{\vc}\rb^2+\O{1}$,
$\ip{\Qca}{\RegSympOp\vc}=\O{1}$, and \eqref{Est:VaRegSympOpVc}
imply that $\lambda_2=\O{\a}$. Equation
\eqref{Eqn:TestFunctionFirstCondition} then implies
$\lambda_3=\O{\a}$.  Evaluating the quadratic form
$\ip{\Lca\cdot}{\cdot}$ at the test function $\e$ and bounding
with H\"{o}lder's inequality implies
\begin{align*}
\ip{\Lca\e}{\e}&\le C\lb\lambda_2^2\HsNorm{1}{\RegSympOp\vc}^2+\lambda_2\lambda_3\HsNorm{1}{\Qca}\HsNorm{1}{\RegSympOp\vc}+\lambda_3^2\HsNorm{1}{\Qca}^2\rb\\
&\le C(c)\a.
\end{align*}
In the last inequality we have used the bounds on $\lambda_1$ and
$\lambda_2$, and the above estimate of
$\HsNorm{1}{\RegSympOp\vc}$.  The constant $C(c)$ does not depend
on the parameter $a$ since $\Hs{1}$ and $\Lp{\infty}$ norms are
translation invariant.

To prove the first part of the proposition, we first prove that
$\inf_X\ip{\Lca\e}{\e}>0$, or equivalently, that
$\inf_{X\cap\Hs{2}}\ip{\Lca\e}{\e}>0$.  By the max-min principle,
$\inf_{X\cap\Hs{2}} \ip{\Lca\e}{\e}$ is attained or is equal to
the bottom of the essential spectrum.  We take $\a$ small enough
so that the above upper bound is below the essential spectrum. Let
$\eta$ be the minimizer.

We claim the set of vectors $\{\va,\, \vc,\ \eta \}$ is a linearly
independent set. If they were dependent, then, since $\va$ and
$\vc$ are orthogonal, there are non-zero constants $\gamma_1$ and
$\gamma_2$ such that $\eta=\gamma_1\va+\gamma_2\vc$. Projecting
this equation onto $\Qca$ and $\RegSympOp\vc$ gives the equations
$\gamma_2\delta'(c)=0$ and
$\gamma_1\ip{\va}{\RegSympOp\vc}+\gamma_2\ip{\RegSympOp\vc}{\va}=0$.
 Thus, if $\a$ is sufficiently small, then both
constants are zero (we have used \eqref{Est:VaRegSympOpVc}).  This
is a contradiction since the zero function does not lie in $X$.

The above argument proves that $\Span{\va,\,\vc,\,\eta}$ is three
dimensional.  The min-max principle states that if
\begin{align*}
E_3&:=\inf_{V\subset\Hs{2},\, \mbox{\scriptsize dim}\,
V=3}\sup_{\e\in V,\,\LpNorm{2}{\e}=1}\ip{\Lca\e}{\e}
\\
&\le \max_{\e\in\,\mbox{\scriptsize
Span}\{\va,\,\vc,\,\eta\},\,\LpNorm{2}{\e}=1 }\ip{\Lca\e}{\e}
\end{align*} is below the essential spectrum, then it is the
third eigenvalue counting multiplicity.  Let
$\e=\gamma_1\eta+\gamma_2\vc+\gamma_3\va$ be the maximizer in the
second line above.  By Proposition \ref{Prop:Spectrum} there are
exactly two non-positive eigenvalues.  Hence
\begin{align*}
0<E_3\le\ip{\Lca\e}{\e}=\gamma_1^2\ip{\Lca\eta}{\eta}-\gamma_2^2\delta'(c).
\end{align*}
Thus, since $\delta'(c)>0$, we must have $\ip{\Lca\eta}{\eta}>0$.
The function $\sigma(c,\a):=\ip{\Lca\eta}{\eta}$ is continuous
with respect to $c$ since both $\Qca$ and $\RegSympOp\vc$ are
continuous as mappings taking $c$ to elements of $\Hs{1}$.  Taking
the infimum of $\sigma(c,\a)$ over $I$ implies $\ip{\Lca\e}{\e}\ge
\rho(\a)\LpNorm{2}{\e}^2$ for all $\e\bot\AnisoRegSympOp\TcaM$,
where $\rho(\a):=\inf_I \sigma(c,\a)$.

To complete the proof of the first statement (modulo the lower
bound on $\varrho(\a)$), we improve the above lower bound to one
involving $\Hs{1}$ norms. If we define
$K:=\sup_I\LpNorm{\infty}{c-f'(\Qca)}$, then
$\ip{\Lca\e}{\e}\ge\LpNorm{2}{\dx\e}^2-K\LpNorm{2}{\e}^2$ for all
$\e\in\Hs{1}$.  Adding a factor $\frac{K+1}{\rho(\a)}$ of this
bound to the above bound gives the required result with
\begin{equation*}
\varrho(\a)=\frac{\rho(\a)}{\rho(\a)+K+1}.
\end{equation*}
Notice that the upper bound $\sigma(c,\a)\le C(c)\a$ derived above
gives, after maximizing constants over $c\in I$, the uniform upper
bound on $\varrho$.

As already shown, the minimizer $\eta$ of $\inf_X\ip{\Lca\e}{\e}$
exists for $\a$ small enough.  We prove the properties of $\eta$
by manipulating its Euler-Lagrange equation
\begin{eqnarray}
\Lca\eta=\beta\eta+\beta_1\Qca+\beta_2\RegSympOp\vc,
\label{Eqn:EulerLagrangeForMinimizer}
\end{eqnarray}
where $\beta$, $\beta_1$, and $\beta_2$ are Lagrange multipliers
for the constraints $\LpNorm{2}{\eta}=1$, $\ip{\eta}{\Qca}=0$, and
$\ip{\eta}{\RegSympOp\vc}=0$.  The inner product of this equation
with $\eta$ shows that $\beta=\sigma$.  Take $\a$ small enough so
that $\beta=\sigma$ is not an eigenvalue of $\Lca$.  Then the
minimizer is unique. Indeed, the difference $\zeta$ between two
minimizers is a solution to $ \Lca\zeta=\beta\zeta$.  Since
$\beta$ is not an eigenvalue, $\zeta=0$ is the only solution to
this equation.

We now decompose $\eta$ orthogonally as $\eta=\gamma\va+\etabot$,
substitute this decomposition into
(\ref{Eqn:EulerLagrangeForMinimizer}), and use $\Lca\va=0$ to
obtain the equation
\begin{eqnarray}
\lb \Lca-\sigma
\rb\etabot=\gamma\sigma\va+\beta_1\Qca+\beta_2\RegSympOp\vc
\label{Eqn:EulerLagrangeForMinimizerDecomposed}
\end{eqnarray}
 for $\etabot$.  To solve for $\etabot$, we first project this equation by
$\NullProjBar$, where $\NullProjBar=1-\NullProj$ and $\NullProj$
is the orthogonal projection onto the nullspace of the operator
$\Lca$:
\begin{eqnarray*}
[L_{Q}-\sigma ]\etabot=\beta_1\Qca+\beta_2 \NullProjBar
\RegSympOp\vc,
\end{eqnarray*}
where $L_Q$ is restriction of $\Lca$ onto the orthogonal
complement of the null space of $\Lca$.  The spectrum of $\Lca$
has essential spectrum $[c,\infty)$ and a finite number of
eigenvalues in $(-\infty,c)$. Moreover, $\Lca$ is independent of
$\a$, and therefore, for $\a$ small enough, we conclude that the
interval $[0,\sigma]$ is disjoint from the spectrum of $L_Q$, with
the distance to the spectrum of $L_Q$ bounded below by a positive
number $C$ independent of $\a$. Hence, we can solve
\eqref{Eqn:EulerLagrangeForMinimizerDecomposed} for $\etabot$ to
obtain
\begin{eqnarray}
\etabot=\lb L_Q-\sigma \rb^{-1}\lb \beta_1\Qca+\beta_2
\NullProjBar\RegSympOp\vc \rb. \label{Eqn:etabot}
\end{eqnarray}

To prove the $\Lp{2}$ the estimate of $\etabot$, we require
estimates on the Lagrange multipliers $\beta_1$ and $\beta_2$.  We
take the inner product of
(\ref{Eqn:EulerLagrangeForMinimizerDecomposed}) with $\va$ and use
\eqref{Est:VaRegSympOpVc} to obtain that
\begin{eqnarray*}
\beta_2[ \delta'(c)+\O{\a} ]=-\gamma\sigma\LpNorm{2}{\va}^2.
\end{eqnarray*}
Thus, $\beta_2=\O{\sigma}$ since the constraint
$\LpNorm{2}{\eta}=1$ implies $\gamma=\O{1}$.  Similarly, since
$\eta$ is orthogonal to $\Qca$, the inner product of
(\ref{Eqn:EulerLagrangeForMinimizer}) with $\vc$ and statement
\ref{LemmaItem:RegularizedSymplecticOperatorInner} of Lemma
\ref{Lemma:RegularizedSymplecticOperatorProperties} gives the
relation
\begin{eqnarray*}
\beta_2\lb \lb\intR{\vc}\rb^2+\O{\a}
\rb=-\sigma\ip{\eta}{\vc}-\beta_1\delta'(c).
\end{eqnarray*}
The estimate $\beta_1=\O{\sigma}$ is immediate using the estimate
of $\beta_2$ and the assumption $\delta'(c)>0$.
 We substitute the estimates of $\beta_1$ and $\beta_2$, and the estimate
$\LpNorm{2}{\RegSympOp\vc}=\mbox{O}(\a^{-\frac{1}{2}})$ into
\eqref{Eqn:etabot}, and use the above fact that $\sigma$ is at
least a distance $C$ away from the spectrum of $L_Q$ to conclude
that
\begin{equation}
\LpNorm{2}{\etabot}=\O{\a^{-\frac{1}{2}}\sigma}.
\label{Est:LTwoNormEtaBot}
\end{equation}
Replacing $\sigma$ with its upper bound gives the third statement
of the proposition.


We now prove a lower bound on the infimum
$\sigma=\inf_X\ip{\Lca\e}{\e}$.  We again need to take $\a$ small
enough so that a minimizer exists.  The orthogonal decomposition
$\eta=\gamma\va+\etabot$ of the minimizer implies
$\sigma=\ip{\Lca\etabot}{\etabot}$. Substituting for $\Lca\etabot$
using \eqref{Eqn:EulerLagrangeForMinimizerDecomposed} gives
\begin{eqnarray*}
\sigma=\sigma\LpNorm{2}{\etabot}^2+\beta_2\ip{\etabot}{\RegSympOp\vc},
\end{eqnarray*}
where we have used that $\ip{\va}{\etabot}=0$ and, since the
minimizer $\eta$ is orthogonal to $\Qca$, $\ip{\etabot}{\Qca}=0$.
Thus, \eqref{Est:LTwoNormEtaBot}, $\beta_2=\O{\sigma}$, and
$\LpNorm{2}{\RegSympOp\vc}=\O{\a^{-1}}$ imply
\begin{eqnarray*}
\sigma\le C_1\frac{\sigma^3}{\a}+C_2\frac{\sigma^2}{\a}
\end{eqnarray*}
or, since $\sigma$ is positive, $C_1\sigma^2+C_2\sigma-\a\ge 0$,
where the constants $C_1$ and $C_2$ depend continuously on $c$.
The positive root of the quadratic is a lower bound on $\sigma$.
After rationalizing, we obtain
\begin{eqnarray*}
\sigma\ge \frac{2\a}{C_2+\sqrt{C_2^2+4C_1\a}}\ge K_1\a,
\end{eqnarray*}
for some constant $K_1$.  Minimizing the constant over $I$
completes the proof of the lower bound.

Our proof of statement three in the proposition requires that
$\inf_Y\ip{\Lca\e}{\e}$ is positive, where $Y:=\{ \e\in \Hs{1}\,
|\,\e\bot\Qc,\va \mbox{and}\ \LpNorm{2}{\e}=1 \}$. The argument is
similar to the proof of $\inf_X\ip{\Lca\e}{\e}$. By the min-max
principle either $\inf_{Y\cap\Hs{2}}=\inf \sigma_{ess}(\Lca)$ or
the minimizer is attained.  There is nothing to prove in the
former case since $\inf \sigma_{ess}(\Lca)=c>0$.  Thus, we assume
$\eta$ is a minimizer. As above, the set $\{\eta,\, \va,\, \vc\}$
is a linearly independent set due to the assumption
$\delta'(c)>0$, and the min-max principle implies the third
eigenvalue $E_3$ satisfies
\begin{eqnarray*}
0<E_3\le \gamma_3^2\ip{\Lca\eta}{\eta}
\end{eqnarray*}
for some constant $\gamma_3$.  Thus, we must have
$\ip{\Lca\e}{\e}>K_3$ for all $\e\in Y$, where $K_3$ is a positive
constant independent of $\a$. As with the infimum over the set
$X$, this inequality can be improved to the $\Hs{1}$ estimate
$\ip{\Lca\e}{\e}>C_3\HsNorm{1}{\e}^2$ for all $\e\bot \Qca,\va$.

We now decompose $\ep$ orthogonally as $\ep=\beta\va+\psi$.  Since
$\ep$ is orthogonal to $\eta$ and $\eta=\gamma\va+\etabot$,
\begin{equation*}
\beta=\LpNorm{2}{\va}^{-2}\gamma^{-1}\ip{\ep}{\eta-\etabot}=\O{\LpNorm{2}{\etabot}\LpNorm{2}{\ep}}=\O{\a^\frac{1}{2}\LpNorm{2}{\ep}}.
\end{equation*}
Substituting this bound into $\HsNorm{1}{\psi}^2\ge
\HsNorm{1}{\ep}^2-\beta^2\HsNorm{1}{\va}^2$ gives that
$\HsNorm{1}{\psi}^2\ge \HsNorm{1}{\ep}^2(1-\a\HsNorm{1}{\va}^2)$.
Thus, if $\a<\frac{1}{2}\HsNorm{1}{\va}^2$, then
$\HsNorm{1}{\psi}^2\ge \frac{1}{2}\HsNorm{1}{\ep}^2$. Substituting
this into the inequality
\begin{equation*}
\ip{\Lca\ep}{\ep}=\ip{\Lca\psi}{\psi}\ge C_3\HsNorm{1}{\ep}^2
\end{equation*}
(which follows from the fact that $\psi\bot$ $\Qca,\va$) completes
the proof.

To prove the last statement we define $\emin:=\e-\ep$.  Since the
vectors $\e$ and $\eta$ are both symplectically orthogonal to the
tangent space, so is $\emin$. Thus, using the above inequalities,
\begin{equation*}
\ip{\Lca\e}{\e}\ge
C_3\HsNorm{1}{\ep}^2+C\a\HsNorm{1}{\emin}^2+2\ip{\e}{\eta}\ip{\Lca\eta}{\ep}.
\end{equation*}
The cross term $\ip{\Lca\eta}{\ep}$ is zero; indeed, substitute
for $\Lca\eta$ using equation
\eqref{Eqn:EulerLagrangeForMinimizer} and use $\ip{\ep}{\eta}=0$
to obtain
\begin{equation*}
\ip{\Lca\eta}{\ep}=\beta_1\ip{\Qca}{\ep}+\beta_2\ip{\RegSympOp\vc}{\ep}.
\end{equation*}
This expression, however, is zero since both $\e$ and $\eta$ are
orthogonal to $\Qca$ and $\RegSympOp\vc$, completing the proof.
\end{proof}


\section{Evolution Equations for the Fluctuation $\e$ and the Parameters $a$ and $c$}
\label{Section:Projection} In Section \ref{Section:Modulation} we
proved that if $u$ remains close enough to the solitary wave
manifold $M_s$, then we can write a solution $u$ to
\eqref{Eqn:KdvGeneralizedWithPotential} uniquely as a sum of a
modulated solitary wave $\Qca$ and a fluctuation $\e$ satisfying
orthogonality condition \eqref{Cond:Orthogonality}. Thus, as $u$
evolves according to the initial value problem
\eqref{Eqn:KdvGeneralizedWithPotential}, the parameters $a(t)$ and
$c(t)$ trace out a path in $\R^2$.  The goal of this section is to
derive the dynamical equations for the parameters $a$ and $c$, and
the fluctuation $\e$. We obtain such equations by substituting the
decomposition $u=\Qca+\e$ into
\eqref{Eqn:KdvGeneralizedWithPotential} and then projecting the
resulting equation onto appropriate directions, with the intent of
using the orthogonality condition on $\e$.

From now on, $u$ is the solution of
\eqref{Eqn:KdvGeneralizedWithPotential} with initial condition
$u_0$ satisfying $\inf_{\Qca\in
M_s}\HsNorm{1}{u_0-\Qca}<<\varepsilon$, and $T_0=T_0(u_0)$ is the
maximal time such that $u(t)\in U_\varepsilon$ for $0\le t\le
T_0$. Then for $0\le t\le T_0$, $u$ can be decomposed as in
\eqref{EquationWithUErrorQDecomposition} and
\eqref{Cond:Orthogonality}.

The majority of the work involves estimating the higher order
terms of the resulting equation for the modulation parameters.  It
turns out that a naive attempt at bounding $\e$ directly with the
Lyapunov method does not give good results.  As will be seen
later, the component of $\e$ in the direction of $\va$ is
particularly problematic.  On the other hand, $\va=-\dx\Qca$ is
the derivative of a function and the null vector of $\Lca$.  This
can be used to improve the bound. Thus, in order to obtain better
estimates on $\e$, we orthogonally decompose the fluctuation as
\begin{equation*}
\e=\emin+\ep
\end{equation*}
where $\emin=\ip{\e}{\eta}\eta$.  Recall that $\eta$ is
approximately $\va$ and is given in Section
\ref{Section:Positivity}: $\eta=\gamma\va+\etabot$, with
$\LpNorm{2}{\eta}=1$, $\gamma=\O{1}$, and
$\LpNorm{2}{\etabot}=\O{\a^\frac{1}{2}}$.  We use the above
decomposition to prove the following proposition regarding the
dynamical equations for $a$ and $c$.

\begin{prop}
\label{Prop:Dynamics} Assume $\delta'(c)>0$ on the compact set
$I\subset \Rp$.  Say $u=\Qca+\e$ is a solution to
\eqref{Eqn:KdvGeneralizedWithPotential}, where $\e$ satisfies
\eqref{Cond:Orthogonality} and $\e=\emin+\ep$ as above.  If
$\a^{-\frac{1}{2}}\HsNorm{1}{\ep}+\HsNorm{1}{\e}$ is small enough
and $\a,\ev\le 1$, then, provided $c\in I$,
\begin{eqnarray}
\label{Eqn:DynamicalEquationForCAndA}
\lb\begin{array}{c} \dot{a}\\
\dot{c}
\end{array}\rb&=&\lb\begin{array}{c}
c-\B(t,a)\\
0
\end{array}\rb+\B'(t,a)\frac{\delta(c)}{\delta'(c)^2}\lb
\begin{array}{c}
-\frac{1}{2}\lb\intR{\vc}\rb^2\\
\delta'(c)
\end{array}\rb+Z(a,c,\e),
\end{eqnarray}
where $|Z(a,c,\e)|\le
C\lb\a\av\ev+\av\ev^2+\lb\a^\frac{1}{2}+\a^{-\frac{1}{2}}\av\ev\rb\HsNorm{1}{\ep}+(\a+\av\ev)\HsNorm{1}{\e}+\HsNorm{1}{\e}^2\rb$,
for some positive constant $C=C(I)$.
\end{prop}
\begin{proof}
Recall that the solitary wave $\Qca$ is an extremal of the
functional $\ActionAtSoliton$. To use this fact we rearrange
definition \eqref{DefinitionLS} of $\ActionAtSoliton$ to write the
Hamiltonian $\HamiltonianWithPotential$ as
\begin{equation*}
\HamiltonianWithPotential(u)=\ActionAtSoliton(u)-cP(u)+\frac{1}{2}\intR{\B
u^2(x)},
\end{equation*}
where for notational simplicity we have suppressed the space and
time dependency of $\B$. Substituting $\Qca+\e$ for $u$ in
\eqref{Eqn:KdVVariationalForm} and using the above expression for
$\HamiltonianWithPotential$ gives the equation
\begin{equation*}
\dot{a}\va+\dot{c}\vc+\dot{\e}=\SympOpInv\ActionAtSoliton'(\Qca+\e)-c\SympOpInv[\Qca+\e]+\SympOpInv[(\Qca+\e)\B],
\end{equation*}
where dots indicate time differentiation.  Taylor expanding
$\ActionAtSoliton'(\Qca+\e)$ to linear order in $\e$ and using
that $\Qca$ is an extremal of $\ActionAtSoliton$ gives
\begin{align}
\dot{\e}=\SympOpInv\left[(\Lca+\DB+\B(a)-c)\e\right]&+\SympOpInv\NpA{\e}-[\dot{a}-c+\B(a)]\va-\dot{c}\vc\nonumber\\&+\B'(a)\SympOpInv[(x-a)\Qca]+\SympOpInv[\RB\Qca].
\label{Eqn:KdVEquationForXiAndParameters}
\end{align}
We have used the relation $\va=-\SympOpInv\Qca$, definition
(\ref{Eqn:Hessian}) of $\Lca$, the definitions
\begin{eqnarray*}
\DB:=\B(x)-\B(a)
\end{eqnarray*}
and
\begin{eqnarray*}
\RB:=\B(x)-\B(a)-\B'(a)(x-a),
\end{eqnarray*}
and definition (\ref{Eqn:NpA}) of $\NpA{\e}$ given in Appendix
\ref{Appendix:EstimateNonlinearRemainders} to write the above
equation in a convenient form. Define the vectors $\zeta_1:=\va$
and $\zeta_2:=\vc$.  Projecting
(\ref{Eqn:KdVEquationForXiAndParameters}) onto
$\AnisoRegSympOp\zeta_i$, for $i=1$ and $2$, and using the
anti-self-adjointness of $\SympOpInv$ gives the two equations
\begin{multline}
[\dot{a}-c+\B(a)]
\left[\ip{\va}{\AnisoRegSympOp\zeta_i}+\ip{\e}{\SympOpInv\AnisoRegSympOp\zeta_i}\right]+\dot{c}\ip{\vc}{\AnisoRegSympOp\zeta_i}+\ip{\dot{\e}}{\AnisoRegSympOp\zeta_i}-\dot{a}\ip{\e}{\SympOpInv\AnisoRegSympOp\zeta_i}=\\
-\B'(a)\ip{(x-a)\Qca}{\SympOpInv\AnisoRegSympOp\zeta_i}-\ip{\RB\Qca}{\SympOpInv\AnisoRegSympOp\zeta_i}\\-\ip{\Lca\e}{\SympOpInv\AnisoRegSympOp\zeta_i}-\ip{\DB\e}{\SympOpInv\AnisoRegSympOp\zeta_i}
-\ip{\NpA{\e}}{\SympOpInv\AnisoRegSympOp\zeta_i}.
\label{Eqn:KdVEquationForXiAndParametersSecond}
\end{multline}
We can replace the term containing $\dot{\e}$ since the time
derivative of the orthogonality condition
$\ip{\e}{\AnisoRegSympOp\zeta_i}=0$ implies
$\langle\dot{\e},\AnisoRegSympOp\zeta_i\rangle=\dot{a}\ip{\e}{\SympOpInv\AnisoRegSympOp\zeta_i}-\dot{c}\ip{\e}{\dc\AnisoRegSympOp\zeta_i}$.
Note that we have used the relation $\p_a\zeta_i=-\dx\zeta_i$.
Thus, equations \eqref{Eqn:KdVEquationForXiAndParametersSecond} in
matrix form are
\begin{align}
(I+B)\AnisoRestNormlzd\lb\begin{array}{c}\dot{a}-c+\B(a)\\
\dot{c}\end{array}\rb=X+Y, \label{Eqn:ApproximateDynamicalSystem}
\end{align}
where
\begin{align*}
X&:=-\B'(a)\delta'(c) \lb
\begin{array}{c}
1\\
0
\end{array}
\rb+\B'(a) \lb
\begin{array}{c}
0\\
\a\ip{(x-a)\Qca}{\RegSympOp\vc}
\end{array}
\rb- \lb
\begin{array}{c}
\ip{\RB\Qca}{\va}\\
\ip{\RB\Qca}{\vc-\a\RegSympOp\vc}
\end{array}
\rb,
\\
Y&:=-\lb
\begin{array}{c}
\ip{\Lca\e+\DB\e+\NpA{\e}}{\va}\\
\ip{\Lca\e+\DB\e+\NpA{\e}}{\vc-\a\RegSympOp\vc}
\end{array}
\rb,
\end{align*}
and
\begin{eqnarray*}
B:=\lb
\begin{array}{cc}
\ip{\e}{\va} & \ip{\e}{\vc}\\
\ip{\e}{\vc} & -\ip{\e}{\dc\RegSympOp\vc}
\end{array}
\rb.
\end{eqnarray*}
We have explicitly computed $\ip{(x-a)\Qca}{\zeta_i}$ and used the
relations $\SympOpInv\AnisoRegSympOp\va=\va$ and
$\SympOpInv\AnisoRegSympOp\vc=\vc-\a\RegSympOp\vc$ to simplify the
above expressions.

We now estimate the error terms and solve for $\dot{a}$ and
$\dot{c}$.  The assumptions we made on the potential imply that
\begin{eqnarray}
|\DB|\le \av\ev(x-a)\ \mbox{and}\ |\RB|\le\av\ev^2(x-a)^2.
\label{Eqn:SizeDV}
\end{eqnarray}
Thus, by H\"{o}lder's inequality and the $\Lp{\infty}$ estimate
$\LpNorm{\infty}{\RegSympOp\zeta_i}\le \LpNorm{1}{\zeta_i}$,
\begin{align*}
\|X\|&=-\B'(a)\delta(c)\lb\begin{array}{c}1\\0\end{array}\rb+\O{\a\av\ev+\av\ev^2}\\
&=\O{\av\ev}.
\end{align*}
In the last equality we have used $\a\le 1$ to bound $\a\av\ev$ by
$\av\ev$.  We now estimate the vector $Y$ using the properties of
$\Lca$ given in Appendix \ref{Prop:PositivityHessian}.  Indeed,
the generalized nullspace relations $\Lca\va=0$ and
$\Lca\vc=\SympOpInv\va$ imply $\ip{\Lca\e}{\zeta_i}=0$, where the
orthogonality condition $\ip{\e}{\AnisoRegSympOp\va}=0$ is used
when $i=2$.  Although the same is not true of the inner product
$\a\ip{\Lca\e}{\RegSympOp\vc}$, we still use the relation
$\Lca\va=0$ to obtain a bound.  If we use
$\eta=\gamma\va+\etabot$, then the orthogonal decomposition
$\emin+\ep$ of $\e$ becomes
\begin{eqnarray*}
\gamma\ip{\e}{\eta}\va+\ep+\ip{\e}{\eta}\etabot,
\end{eqnarray*}
where $\gamma=\O{1}$, $\LpNorm{2}{\etabot}=\O{\a^\frac{1}{2}}$ and
$\LpNorm{2}{\eta}=1$. Thus,
\begin{align*}
\a\ip{\Lca\e}{\RegSympOp\vc}&=\a\ip{\Lca\ep}{\RegSympOp\vc}+\a\ip{\e}{\eta}\ip{\Lca\etabot}{\RegSympOp\vc}\\
&=\O{\a^\frac{1}{2}\HsNorm{1}{\ep}+\a\HsNorm{1}{\e}}.
\end{align*}
The terms containing the potential, but not the term
$\RegSympOp\vc$, are easily estimated using the bound on $\DB$ and
exponential decay of $\va$ and $\vc$.  The resulting estimates are
\begin{equation*}
\ip{\DB\e}{\va}=\O{\av\ev\LpNorm{2}{\e}}\ \mbox{and}\
\ip{\DB\e}{\vc}=\O{\av\ev\LpNorm{2}{\e}}.
\end{equation*}
To obtain a bound of $\a\ip{\DB\e}{\RegSympOp\vc}$, we decompose
$\e$ as above to obtain
\begin{eqnarray*}
\a\ip{\DB\e}{\RegSympOp\vc}=\gamma\a\ip{\e}{\eta}\ip{\va}{\DB\RegSympOp\vc}+\a\ip{\e}{\eta}\ip{\etabot}{\DB\RegSympOp\vc}+\a\ip{\ep}{\DB\RegSympOp\vc}.
\end{eqnarray*}
Then the estimates $\SupNorm{\RegSympOp\vc}=\O{1}$ and
$\LpNorm{2}{x\RegSympOp\vc}=\O{\a^{-\frac{3}{2}}}$ imply
\begin{eqnarray*}
\a\ip{\DB\e}{\RegSympOp\vc}=\O{\a^{-\frac{1}{2}}\av\ev\HsNorm{1}{\ep}+\av\ev\HsNorm{1}{\e}}.
\end{eqnarray*}
Lastly, H\"{o}lder's inequality and Lemma
\ref{Lemma:NonlinearEstimates} imply
$\ip{\NpA{\e}}{\zeta_i}=\O{\HsNorm{1}{\e}^2}$ and
$\a\ip{\NpA{\e}}{\RegSympOp\vc}=\O{\a^\frac{1}{2}\HsNorm{1}{\e}^2}$.
Adding the above estimates gives
\begin{eqnarray*}
\| Y
\|=\lb\a^\frac{1}{2}+\a^{-\frac{1}{2}}\av\ev\rb\HsNorm{1}{\ep}+\lb
\a+\av\ev \rb\HsNorm{1}{\e}+\HsNorm{1}{\e}^2.
\end{eqnarray*}
The inner products $\ip{\e}{\va}$ and $\ip{\e}{\vc}$ are clearly
of order $\HsNorm{1}{\e}$.  We estimate the remaining entry
$\ip{\e}{\dc\RegSympOp\vc}=\ip{\e}{\RegSympOp\dc^2\Qca}$ of $B$ by
the same technique as above: we replace $\e$ by
$\gamma\ip{\e}{\eta}\va+\ep+\ip{\e}{\eta}\etabot$ and estimate the
resulting expression to obtain
\begin{equation}
\|B\|=\O{\a^{-\frac{1}{2}}\HsNorm{1}{\ep}+\HsNorm{1}{\e}}.
\end{equation}
We take $\|B\|$ smaller than one so that $I+B$ is invertible and
$\|(I+B)^{-1}\|=\O{1}$. Acting on
\eqref{Eqn:ApproximateDynamicalSystem} by
$(I+B)^{-1}=I-B(I+B)^{-1}$ and then $\AnisoRestNormlzd$ gives that
\begin{eqnarray*}
\lb
\begin{array}{c}
\dot{a}-c+\B(a)\\
\dot{c}
\end{array}
\rb= \AnisoRestNormlzd X-\AnisoRestNormlzd
B(I+B)^{-1}X+\AnisoRestNormlzd(I+B)^{-1} Y.
\end{eqnarray*}
We use the leading order expressions for $\AnisoRestNormlzd$ and
$X$, and the bounds on $\|X\|$, $\|Y\|$, $\|B\|$ and
$\|(I+B)^{-1}\|$ to obtain the estimate
\begin{eqnarray*}
\lb
\begin{array}{c}
\dot{a}-c+\B(a)\\
\dot{c}
\end{array}
\rb&=&\B'(a)\frac{\delta(c)}{\delta'(c)^2}\lb\begin{array}{c}-\frac{1}{2}\lb\intR{\vc}\rb^2\\
\delta'(c)
\end{array}\rb\\
&&+\O{\a\av\ev+\av\ev^2+\lb\a^\frac{1}{2}+\a^{-\frac{1}{2}}\av\ev\rb\HsNorm{1}{\ep}+(\a+\av\ev)\HsNorm{1}{\e}+\HsNorm{1}{\e}^2}.
\end{eqnarray*}

In the order notation used above, the implicit constants are
continuous with respect to the parameter $c$ and independent of
the parameter $a$.  Maximizing these constants over the compact
set $I$ completes the proof.
\end{proof}


\section{The Lyapunov Function}
\label{Section:LyapDeriv} In the last section we derived dynamical
equations for the modulation parameters.  These equations contain
the $\Hs{1}$ norm of the fluctuation.  In this section we begin to
prove a bound on $\e$.  Recall that the latter bound is needed to
ensure that $u$ remains close to the manifold of solitary waves
$M_s$ for long time.

We employ a Lyapunov argument with Lyapunov function
\begin{align}
\LyapunovFunctional(t):=\ActionAtSoliton(\Qca+\e)-\ActionAtSoliton(\Qca)+\B'(a)\ip{(x-a)\Qca}{\e}.
\label{equ:LSDiffDef}
\end{align}
Remark: if $f(u)=u^3$, the last term in the Lyapunov functional is
not needed; however, apart from computational complexity, there is
no disadvantage in using the above function for this special case
as well.

\begin{lemma}
\label{Lemma:AlmostConservationOfLyapunov} Assume $\delta'(c)> 0$
on the set $I$.  Let $u=\Qca+\e$ be as in Proposition
\ref{Prop:Dynamics}. Let $\a, \av$, $\ev$, $\HsNorm{1}{\e}\le 1$.
Then, if $\a^{-\frac{1}{2}}\HsNorm{1}{\ep}+\HsNorm{1}{\e}$ is
small enough and $c\in I$, there is a constant $C=C(I)$ such that
\begin{align}
\dd{t} \LyapunovFunctional(t)&\le C[(\av\ev+\HsNorm{1}{\e}^2)\Xi],
\label{TimeDerivativeLiapunovFunctional}
\end{align}
where
$\Xi:=\av\ev+\lb\a^\frac{1}{2}+\a^{-\frac{1}{2}}\av\ev\rb\HsNorm{1}{\ep}+(\a+\ev+\tv)\HsNorm{1}{\e}+\HsNorm{1}{\e}^2$.
\end{lemma}
\begin{proof}
Suppressing explicit dependence on $x$ and $t$, we have by
definition
\begin{align*}
\ActionAtSoliton(u):=\HamiltonianWithPotential(u)-\frac{1}{2}\intR{
u^2 \B}+cP(u).
\end{align*}
Thus, relations (\ref{Eqn:ConservationHamiltonian}),
(\ref{Eqn:ConservationMomentum}) and
(\ref{Eqn:ConservationPotentialMomentum}) imply that the time
derivative of $\ActionAtSoliton$ along the solution $u$ is
\begin{align*}
\dd{t}\ActionAtSoliton(u)=\intR{\frac{1}{2}\dot{c} u^2
+\B'\left[\frac{1}{2}c u^2- u f(u)+\frac{3}{2}(\dx
u)^2+F(u)\right]+\B''\, u \dx u}.
\end{align*}
Substituting $\Qca+\e$ for $u$, manipulating the result using
antisymmetry of $\dx$, and collecting appropriate terms into
$\B'(a)\ip{\Lca\e}{\SympOpInv((x-a)\Qca)}$,
$\ip{\NpA{\e}}{\dx[\DB(\Qca+\e)]}$, and
$\ip{\ActionAtSoliton'(\Qca)}{\dx(\DB(\Qca+\e))}$ gives the
relation
\begin{align*}
\dd{t}[\ActionAtSoliton(\Qca+\e)-\ActionAtSoliton(\Qca)]=&\B'(a)\ip{\Lca\e}{\SympOpInv((x-a)\Qca)}+\dot{c}\ip{\Qca}{\e}+\ip{\Lca\e}{\dx\lb\RB\Qca\rb}+\dot{c}\frac{1}{2}\LpNorm{2}{\e}^2\\
&+c\frac{1}{2}\ip{\B'\e}{\e}+\frac{3}{2}\ip{\B'\dx\e}{\dx\e}-\ip{f'(\Qca)\e}{\dx(\DB\e)}\\
&+\ip{\NpA{\e}}{\dx[\DB(\Qca+\e)]}+\ip{\B''\e}{\dx\e}+\ip{\ActionAtSoliton'(\Qca)}{\dx[\DB(\Qca+\e)]}.
\end{align*}
The last term is zero because $\ActionAtSoliton'(\Qca)=0$.  The
inner product $\ip{\e}{\Qc}$ is equal to
$-\a\ip{\e}{\RegSympOp\Qca}=\O{\a^\frac{1}{2}\HsNorm{1}{\e}}$
since $-\RegSympOp\va=\Qca-\a\RegSympOp\Qca$ and $\e\bot\Qca$. We
use Lemma \ref{Lemma:NonlinearEstimates}, assumptions
(\ref{Eqn:AssumptionOnPotential}) on the potential, estimates
(\ref{Eqn:SizeDV}), and
\begin{align*}
|\delta \B'|&\le\av\ev^2 x\
\end{align*}
to estimate the size of the time derivative.  We also use that
$\Qca$, $\dx\Qca$, $\dx^2\Qca$ and $f'(\Qca)$ are exponentially
decaying.  When $\ev\le 1$, higher order terms like
$\ip{\B''\e}{\dx\e}$ are bounded above by lower order terms like
$\ip{\B'\e}{\e}$.  Similarly, if $\HsNorm{1}{\e}\le 1$, then
$\av\ev\HsNorm{1}{\e}^2\le\av\ev\HsNorm{1}{\e}$.  This procedure
gives the estimate
\begin{align*}
\dd{t}[\ActionAtSoliton(\Qca+\e)-\ActionAtSoliton(\Qca)]=&\B'(a)\ip{\e}{\Lca\SympOpInv((x-a)\Qca)}+\ip{\NpA{\e}}{\DB\dx\e}\\
&+\O{|\dot{c}|\HsNorm{1}{\e}^2+\av\ev^2\HsNorm{1}{\e}+\av\ev\HsNorm{1}{\e}^2}.
\end{align*}
We compute
\begin{align*}
\ip{\NpA{\e}}{\DB\dx\e}=&\ip{\NpA{\e}+\frac{1}{2}f''(\Qc)\e^2}{\DB\dx\Qc}\\
&-\intR{\lb
F(\Qca+\e)-F(\Qca)-f(\Qca)\e-\frac{1}{2}f'(\Qca)\e^2\rb \B'},
\end{align*}
and use the second estimate and the proof of the third estimate of
 Lemma \ref{Lemma:NonlinearEstimates} to obtain
$\ip{\NpA{\e}}{\DB\dx\e}=\O{\av\ev\HsNorm{1}{\e}^3}$. Thus, since
$\av\ev\HsNorm{1}{\e}^3\le \av\ev\HsNorm{1}{\e}^2$ when
$\HsNorm{1}{\e}\le 1$, we have
\begin{multline}
\dd{t}[\ActionAtSoliton(\Qca+\e)-\ActionAtSoliton(\Qca)]=\B'(a)\ip{\e}{\Lca\SympOpInv((x-a)\Qca)}\\
+\O{|\dot{c}|\HsNorm{1}{\e}^2+\av\ev^2\HsNorm{1}{\e}+\av\ev\HsNorm{1}{\e}^2}.
 \label{Eqn:PropAlmostLiapunovConservationDLS}
\end{multline}
When $f(u)=u^3$, $\ip{\e}{\Lca\SympOpInv((x-a)\Qca)}=0$ since
$\vc=\SympOpInv[(x-a)\Qca]$.  In this special case the above
estimate is sufficient for our purposes, but in general, we need
to use the corrected Lyapunov functional. When $\e\in
C(\R,\,\Hs{1})\cap C^1(\R,\, \Hs{-2})$, $\B'(a)\ip{\e}{(x-a)\Qca}$
is continuously differentiable with respect to time;
\begin{align*}
\dd{t}\left[ \B'(a)\ip{\e}{(x-a)\Qca}
\right]=&\dt\B'\ip{\e}{(x-a)\Qca}+\B'(a)\ip{\dot{\e}}{(x-a)\Qca}+\dot{c}\B'(a)\ip{\e}{(x-a)\vc}\\
&+\dot{a}\B'(a)\ip{\e}{(x-a)\va}+\dot{a}\B''(a)\ip{\e}{(x-a)\Qca},
\end{align*}
where $\ip{\e}{\Qca}=0$ has been used to simplify the derivative.
Substituting for $\dt\e$ using
(\ref{Eqn:KdVEquationForXiAndParameters}) gives
\begin{align*}
\dd{t}[\B'(a)\ip{\e}{(x-a)\Qca}]=&-\B'(a)\ip{\e}{\Lca\SympOpInv((x-a)\Qca)}-[\dot{a}-c+\B(a)]\B'(a)\frac{1}{2}\LpNorm{2}{\Qca}^2+\dt\B'\ip{\e}{(x-a)\Qca}\\
&+[\dot{a}-c+\B(a)]\B'(a)\ip{\dx\e}{(x-a)\Qca}+[\dot{a}-c+\B(a)] \B''(a)\ip{\e}{(x-a)\Qca}\\
&+\dot{c} \B'(a)\ip{\e}{(x-a)\vc}-\B'(a)\ip{\e}{\DB\dx((x-a)\Qca)}-\B'(a)\ip{\NpA{\e}}{\dx((x-a)\Qca)}\\
&-\B'(a)\ip{\RB\Qca}{\dx((x-a)\Qca)}+[c-\B(a)]\B''(a)\ip{\e}{(x-a)\Qca}.
\end{align*}
We estimate using the same assumptions used to derive
(\ref{Eqn:PropAlmostLiapunovConservationDLS}).  If
$\HsNorm{1}{\e}$ and $\ev$ are less than 1, then
\begin{align*}
\dd{t}[\B'(a)\ip{\e}{(x-a)\Qca}]=&-\B'(a)\ip{\e}{\Lca\SympOpInv((x-a)\Qca)}+\O{ |\dot{a}-c+\B(a)|\av\ev+|\dot{c}|\av\ev\HsNorm{1}{\e}}\\
&+\O{\av^2\ev^3+((1+\av)\ev^2+\ev\tv)\av\HsNorm{1}{\e}+\av\ev\HsNorm{1}{\e}^2}.
\end{align*}
Adding the above expression to
(\ref{Eqn:PropAlmostLiapunovConservationDLS}) gives an upper bound
containing $|\dot{c}|$ and $|\dot{a}-c+\B(a)|$. Replacing these
quantities using the bound
\begin{eqnarray*}
|\dot{c}|+|\dot{a}-c+\B(a)|=\O{\av\ev+\lb\a^\frac{1}{2}+\a^{-\frac{1}{2}}\av\ev\rb\HsNorm{1}{\ep}+(\a+\av\ev)\HsNorm{1}{\e}+\HsNorm{1}{\e}^2}
\end{eqnarray*}
from Proposition \ref{Prop:Dynamics}, and bounding higher order
terms by lower order terms gives
(\ref{TimeDerivativeLiapunovFunctional}). To use the above bounds
on $|\dot{c}|$ and $|\dot{a}-c+\B(a)|$ we must assume
$\a^{-\frac{1}{2}}\HsNorm{1}{\ep}+\HsNorm{1}{\e}$ is small enough
so that Proposition \ref{Prop:Dynamics} holds.
\end{proof}

\section{Bound on the Fluctuation and Proof of Main Theorem}
\label{Section:BoundOnFluct} We are now in a position to prove the
bound on $\e$.
\begin{prop}
\label{Prop:BoundOnFluctuation} Say $u=\Qca+\e$ is a solution to
(\ref{Eqn:KdvGeneralizedWithPotential}), where $\e$ satisfies
(\ref{Cond:Orthogonality}).  Let $\av,\,\ev\le 1$ and
$0<s<\frac{1}{2}$.  Then, if $\av\ev$ is small enough, there are
constants $C_1$, $C_2$, and $C_3$, such that if the initial
condition $u_0$ satisfies $\inf_{\Qca\in
M_s}\HsNorm{1}{u_0-\Qca}<<(\av\ev)^{2 s}$, then
\begin{eqnarray*}
\HsNorm{1}{\e(t)}\le C_1(\av\ev)^s\ \mbox{and}\
\HsNorm{1}{\ep(t)}\le C_2(\av\ev)^\frac{3 s}{2},
\end{eqnarray*}
for all times $t\le T_1:=C_3\lb(\av\ev)^s+\tv+\ev\rb^{-1}$.
\end{prop}
\begin{proof}
We choose $\epsilon_0:=\inf_{\Qca\in M_s}\HsNorm{1}{u_0-\Qca}$
small enough so that $\HsNorm{1}{\e(0)}\le
C\a^{-\frac{1}{2}}\epsilon_0< \varepsilon$ (see
\eqref{Ineq:InitialConditionIFT}) is small enough to satisfy the
conditions of Proposition \ref{Prop:ExistenceOfDecomposition} and
Lemma \ref{Lemma:AlmostConservationOfLyapunov}.  Then, continuity
of the solution $u=\Qca+\e$ in $\Hs{1}$ with respect to time
implies the conditions continue to be satisfied over a non-empty
time interval $[0,T]$. We will obtain an estimate of
$\HsNorm{1}{\e(t)}$ over a time interval $[0,T]$ by deriving an
equality for $\HsNorm{1}{\e(t)}$ from upper and lower bounds on
the Lyapunov functional.  We suppress dependence on $t$ for
notational convenience.

Define $\HTNorm{\e}:=\sup_{[0,T]}\HsNorm{1}{\e(t)}$ and
$|\Xi|_T:=\sup_{[0,T]}|\Xi|$.  Integrating the time maximized
upper bound in Lemma \ref{Lemma:AlmostConservationOfLyapunov}
gives
\begin{equation*}
\LyapunovFunctional(t)\le
|\LyapunovFunctional(0)|+C(\av\ev+\HTNorm{\e}^2)|\Xi|_T T
\end{equation*}
for all $t\in[0,T]$.  A lower bound is obtained by expanding the
$\ActionAtSoliton(\Qca+\e)$ term in $\LyapunovFunctional(t)$ to
quadratic order and using $\ActionAtSoliton'(\Qca)=0$ to obtain
\begin{equation*}
\LyapunovFunctional(t)=\ip{\Lca\e}{\e}+\NA{\e}+\B'(a)\ip{\e}{(x-a)\Qca},
\end{equation*}
where the nonlinear remainder $\NA{\e}$ is defined in Appendix
\ref{Appendix:EstimateNonlinearRemainders}.  Estimating $\NA{\e}$
with Lemma \ref{Lemma:NonlinearEstimates} and using anisotropic
coercivity of the Hessian $\Lca$ (Proposition
\ref{Prop:PositivityHessian}) gives
\begin{equation*}
\LyapunovFunctional(t)\ge
C_3\HsNorm{1}{\ep}^2+C\a\HsNorm{1}{\emin}^2-C(\av\ev\HsNorm{1}{\e}+\HsNorm{1}{\e}^3).
\end{equation*}
Together with the upper bound on $\LyapunovFunctional(t)$, this
bound implies
\begin{equation}
\HsNorm{1}{\ep}^2,\,\a\HsNorm{1}{\emin}^2\le\HsNorm{1}{\ep}^2+\a\HsNorm{1}{\emin}^2\le
|\LyapunovFunctional(0)|+(\av\ev+\HTNorm{\e}^2)|\Xi|_T
T+\av\ev\HTNorm{\e}+\HTNorm{\e}^3.
\label{Est:LowerBoundXiPerpXiMin}
\end{equation}
Note that we have set non-essential constants to unity.  Since the
above inequalities hold for all $t\in[0,T]$, we can replace
$\HsNorm{1}{\emin}^2$ with $\HTNorm{\emin}^2$ and
$\HsNorm{1}{\ep}^2$ with $\HTNorm{\ep}^2$.  Multiplying the
resulting inequality for $\HTNorm{\ep}^2$ by $\a$ and adding to
the inequality for $\a\HsNorm{1}{\emin}^2$ gives
\begin{equation*}
\a\HTNorm{\e}^2\le
|\LyapunovFunctional(0)|+(\av\ev+\HTNorm{\e}^2)|\Xi|_T
T+\av\ev\HTNorm{\e}+\HTNorm{\e}^3.
\end{equation*}
Next, we take $|\Xi|_T T=\O{\a}$ and $\HTNorm{\e}=\O{\a}$ to
obtain the bound
\begin{equation*}
\HTNorm{\e}\le
\a^{-\frac{1}{2}}|\LyapunovFunctional(0)|^\frac{1}{2}+(\av\ev)^\frac{1}{2}+\av\ev\a^{-1}.
\end{equation*}

The initial value of the Lyapunov functional
$\LyapunovFunctional(0)$ can be bounded by the $\Hs{1}$ norm of
the initial fluctuation $\HsNorm{1}{\e(0)}\le
C\a^{-\frac{1}{2}}\epsilon_0$.  Indeed, Taylor expanding
$\ActionAtSoliton(\Qca+\e)$ to second order in $\e$, and using the
third estimate in Lemma \ref{Lemma:NonlinearEstimates} gives
$|M(0)|=\O{\a^{-1}\epsilon_0^2+\a^{-\frac{1}{2}}\av\ev\epsilon_0}$
if $\epsilon_0\le 1$.

By order considerations, if we choose $\a=(\av\ev)^s$, with
$0<s<\frac{1}{2}$, and if $\epsilon_0<<(\av\ev)^{2 s}$, then the
bound $\HTNorm{\e}=\O{(\av\ev)^s}$ holds for $T=\O{\a|\Xi|_T}$.
The bound $\HTNorm{\ep}=\O{(\av\ev)^\frac{3s}{2}}$ is obtained by
substituting the bound for $\HsNorm{1}{\e}$ into second inequality
of \eqref{Est:LowerBoundXiPerpXiMin}.  We obtain the conservative
estimate $T_1=\O{[(\av\ev)^s+\tv+\ev]^{-1}}$ by substituting the
bounds on $\HTNorm{\e}$ and $\HTNorm{\ep}$ into the expression for
$|\Xi|_T$.  To complete the proof, we take $\epsilon_0$, $\av$,
and $\ev$ sufficiently small so that the smallness assumptions in
all the propositions and lemmas hold.
\end{proof}

We now prove the main theorem.
\begin{proof}[Proof of Theorem \ref{MainThm}]
By our choice $\epsilon_0<\varepsilon$, there is a (maximal) time
$T_0$ such that the solution $u$ in
\eqref{Eqn:KdvGeneralizedWithPotential} is in $U_\varepsilon$ for
time $t\le T_0$.  Hence decomposition
\eqref{EquationWithUErrorQDecomposition} with
\eqref{Cond:Orthogonality} and Proposition
\ref{Prop:BoundOnFluctuation} are valid for $u$ and imply the
statements of the main theorem and in particular
$\HsNorm{1}{\e(t)}\le C_1(\av\ev)^s$ for times $t\le \min\{T_0,
T_1\}$.  If we take $\av\ev$ such that
$C_1(\av\ev)^s<\varepsilon=\O{(\av\ev)^\frac{s}{2}}$, then the
above bound holds for $t\le T_1$ by maximality of the time $T_0$.
\end{proof}

\appendix
\appendixpage
\section{Global Wellposedness of the bKdV}
\label{Appendix:GlobalWellposedness} \indent\indent In this
appendix we prove Theorem \ref{Thm:GWP}, global wellposedness of
the bKdV equation
\begin{equation}
\dt u=-\dx(\dx^2 u+u^2+\B(t,x) u) \label{Eqn:AppendixbKdV}
\end{equation}
in $\Hs{1}$, with an appropriate condition on $\B$.  We extend the
local wellposedness proof of Kenig-Ponce-Vega \cite{KePoVe1993} in
the case of $b=0$, and use an energy argument to extend local
wellposedness to global wellposedness.  Define
\begin{equation*}
\LXLTNorm{p}{q}{f}:=\lb\intR{\lb\int_{-T}^T f(x,t)^q}\,
dt\rb^\frac{p}{q}\rb^\frac{1}{p}
\end{equation*}
and similarly the $L^p_T L^q_X$, $L_T^p H^1_X$ and $L_T^p
W_X^{k,p}$ norms (recall $W^{k,p}$ is the Sobolev space based on
$\Lp{p}$).  Let $\hat{f}$ denote the Fourier transform of $f$.

We begin with a lemma.
\begin{lemma}
\label{Lemma:CommEstimate} Let $g\in C(\C)$ and $g'\in
L^\infty(\C)$ (in distribution). If $\B\in\Lp{2}$ with
$\widehat{\B'}\in\Lp{1}$, then
\begin{equation*}
\left\|\comm{g(\dx)}{b}\right\|_{\Lp{2}\rightarrow\Lp{2}}\le
C\LpNorm{\infty}{g'}\LpNorm{1}{\widehat{\B'}}.
\end{equation*}
\end{lemma}
\begin{proof}
Writting $\B$ as a Fourier integral and interchanging the integral
and commutator gives
\begin{equation}
\comm{g(\dx)}{\B}=\frac{1}{(2\pi)^\frac{1}{2}}\int_{-\infty}^\infty
\hat{b}(k)\comm{g(\dx)}{e^{i k x}}\ dk.
\label{Eqn:AppendixBFourierIntegral}
\end{equation}
We require an estimate of $\comm{g(\dx)}{e^{i k x}}=e^{i k x}\lb
e^{-i k x}g(\dx) e^{i k x}-g(\dx)\rb$.  The term enclosed in
brackets can be written as an integral:
\begin{equation*}
\comm{g(\dx)}{e^{i k x}}=e^{i k x}\int_0^1 \frac{d}{ds}\lb e^{-i k
x s}g(\dx) e^{i k x s} \rb\ ds,
\end{equation*}
or on differentiating and computing in Fourier space,
\begin{equation*}
\comm{g(\dx)}{e^{i k x}}=i k e^{i k x}\int_0^1 g'(\dx+i k s)\ ds.
\end{equation*}
Hence $\LpNorm{2}{\comm{g(\dx)}{e^{i k x}} f}\le
|k|\LpNorm{\infty}{g'}\LpNorm{2}{f}$ for all $f\in\Lp{2}$.
Substituting this bound into \eqref{Eqn:AppendixBFourierIntegral}
proves the lemma.
\end{proof}

\begin{proof}[Proof of Theorem \ref{Thm:GWP}]
The proof is $p$ power specific ($f=u^p$); we only present the
proof for p=2.  We begin by proving local wellposedness. Let $W$
be the unitary group generated by $-\dx^3$; that is,
$W(t):=e^{-\dx^3 t}$. Consider the bKdV without the term $\B u$.
We rewrite this equation as a fixed point problem $u=\Phi(u)$,
where the map $\Phi$ is defined by
\begin{equation*}
\Phi:v\mapsto W(t)u_0-\int_0^t W(t-\tau)v \dx v\, d\tau.
\end{equation*}
Given an initial condition $u_0$, Kenig, Ponce and Vega
\cite{KePoVe1993} proved that $\Phi$ is a strict contraction on a
ball $B_{X_T^s}(a)$ of radius $a=a(\HsNorm{s}{u_0})$ and centre
$v=0$ in the space $X^s_T$ for all $s>\frac{3}{4}$, where
\begin{equation*}
X_T^s:=\{ v\in C([-T,T],\Hs{s})\, |\, \Lambda^T(v)\le \infty \},
\end{equation*}
and
\begin{align*}
\Lambda_T^s(v):=\LTHXNorm{\infty}{s}{v}+\LTLXNorm{4}{\infty}{\dx
v}+\LXLTNorm{\infty}{2}{\Dx^s\dx
v}+(1+T)^{-\rho}\LXLTNorm{2}{\infty}{v}.
\end{align*}
Here $\rho>\frac{3}{2}$. More precisely, given $\HsNorm{s}{u_0}$
and $0<\epsilon\le 1$, there is an $a$ and $T$ such that $u_0\in
B_{X_T^s}(a)$, $\Phi:B_{X_T^s}(a)\rightarrow B_{X_T^s}(\epsilon
a)$ and $\Lambda_T^s(\Phi(v)-\Phi(\tilde{v}))\le
\epsilon\Lambda_T^s(v-\tilde{v})$.

We now formulate \eqref{Eqn:AppendixbKdV} as a fixed point
problem.  Define
\begin{equation*}
\Psi:v\mapsto -\int_0^t W(t-\tau)\dx(\B v)\, d\tau.
\end{equation*}
Then, the bKdV is equivalent to solving $u=\Phi(u)+\Psi(u)$. Thus,
if $\Phi+\Psi$ is a strict contraction on some ball
$B_{X_T^s}(a)$, then equation \eqref{Eqn:AppendixbKdV} has a
solution in the same class as \eqref{Eqn:AppendixbKdV} with
$\B=0$.  To prove this we need estimates of $\Lambda_T^s(\Psi(v))$
and $\Lambda_T^s(\Psi(v)-\Psi(\tilde{v}))$ for $v\in
B_{X_T^s}(a)$.  We will make use of the estimates
\begin{align}
\LTLXNorm{4}{\infty}{\Dx^\frac{1}{4} W(t) g(x)}&\le C\LpXNorm{2}{g},\label{Ineq:KePoVeTwo}\\
\LXLTNorm{\infty}{2}{\dx W(t) g(x)}&\le C\LpXNorm{2}{g},\label{Ineq:KePoVeThree}\\
\LXLTNorm{2}{\infty}{W(t)g(x)}&\le
C(1+T)^\rho\HsXNorm{1}{g}\label{Ineq:KePoVeFour},
\end{align}
the proofs of which are given in \cite{KePoVe1993}.

We estimate $\Lambda_T^s(\Psi(v))$ for $s=1$, beginning with
$\LTHXNorm{\infty}{1}{\Psi(v)}$. The inequality
\begin{align*}
\LTHXNorm{\infty}{1}{\int_0^t W(t-\tau)\dx(\B v)\,d\tau}&\le
C\LTLXNorm{\infty}{2}{\int_0^t W(t-\tau)\dx(\B v)\,
d\tau}\\&+C\LTLXNorm{\infty}{2}{\int_0^t W(t-\tau)\Dx\dx(\B
v)\,d\tau}
\end{align*}
follows from the commutativity of $\Dx=-i\dx$ with $W(t-\tau)$.
Taking the $L_X^2$ norm inside the integrals via Minkowski's
inequality, and using that $W(t-\tau)$ is unitary gives that
\begin{equation*}
\LTHXNorm{\infty}{1}{\Psi(v)}\le C\LpTNorm{\infty}{\int_0^t
\LpXNorm{2}{\dx(\B v)}+\LpXNorm{2}{\Dx\dx(\B v)}\, d\tau}.
\end{equation*}
We increase the integration domain to $[-T,T]$ and use
H\"{o}lder's inequality to obtain the estimate
\begin{equation*}
\LTHXNorm{\infty}{1}{\Psi(v)}\le C\lb \LTLXNorm{2}{2}{\dx(b
v)}+\LTLXNorm{2}{2}{\Dx\dx(b v)} \rb.
\end{equation*}

The bound
\begin{equation*}
\LTLXNorm{4}{\infty}{\dx\Psi(v)}\le
\LpTNorm{4}{\int_{-T}^T\LpXNorm{\infty}{\dx W(t-\tau)\dx(b v)}\
d\tau}
\end{equation*}
is obtained from $\LTLXNorm{4}{\infty}{\dx\Psi(v)}$ by moving the
derivative $\dx$ and $\Lp{\infty}$ norm into the integral over
$\tau$, and increasing the domain of integration to $[-T,T]$.
Minkowski's inequality then implies
\begin{equation*}
\LTLXNorm{4}{\infty}{\dx\Psi(v)}\le
\int_{-T}^T\LTLXNorm{4}{\infty}{\dx W(t-\tau)\dx(b v)}\ d\tau.
\end{equation*}
We use that $\dx$ commutes with $W$ and the group properties of
$W$ to rewrite this inequality:
\begin{equation*}
\LTLXNorm{4}{\infty}{\dx\Psi(v)}\le
\int_{-T}^T\LTLXNorm{4}{\infty}{\Dx^\frac{1}{4}
W(t)W(-\tau)\sigma\Dx^\frac{3}{4}\dx(b v)}\ d\tau,
\end{equation*}
where $\sigma$ is multiplication by $i\, sgn(k)$ in Fourier space.
The quantity $W(-\tau)\sigma\Dx^\frac{3}{4}\dx(\B v)$ does not
depend on time $t$.  Thus, we use estimate \eqref{Ineq:KePoVeTwo}
and that $W$ and $\sigma$ preserve the $\Lp{2}$ norm to obtain the
bound
\begin{align*}
\LTLXNorm{4}{\infty}{\dx\Psi(v)}&\le
\int_{-T}^T\LpXNorm{2}{\Dx^\frac{3}{4}\dx(\B v)}\, d\tau
\end{align*}
or $\LTLXNorm{4}{\infty}{\dx\Psi(v)}\le C\lb
\LTLXNorm{2}{2}{\dx(\B v)}+\LTLXNorm{2}{2}{\Dx\dx(\B v)}\rb$.

Again, Minkowski's inequality and the properties of $W$ give the
bound
\begin{equation*}
\LXLTNorm{\infty}{2}{\Dx\dx v}\le\int_0^t\LXLTNorm{\infty}{2}{\Dx
W(t)W(-\tau)\dx^2(\B v)}\ d\tau.
\end{equation*}
Since the same holds with integration over $[-T,T]$ and
$W(-\tau)\dx^2(b v)$ is independent of time $t$,
\eqref{Ineq:KePoVeThree} implies
\begin{equation*}
\LXLTNorm{\infty}{2}{\Dx\dx v}\le \LTLXNorm{1}{2}{W(-\tau)\Dx\dx(b
v)}\le\LTLXNorm{2}{2}{\Dx\dx(b v)}.
\end{equation*}

As above, we find that
\begin{equation*}
(1+T)^{-\rho}\LXLTNorm{2}{\infty}{\Psi(v)}\le
(1+T)^{-\rho}\int_{-T}^T\LXLTNorm{2}{\infty}{W(t)W(-\tau)\dx(\B
v)}\ d\tau.
\end{equation*}
Estimate \eqref{Ineq:KePoVeFour} then implies
\begin{equation*}
(1+T)^{-\rho}\LXLTNorm{2}{\infty}{\Psi(v)}\le
\LTHXNorm{1}{1}{W(-\tau)\dx(\B v)}\le C\lb \LTLXNorm{2}{2}{\dx(\B
v)}+\LTLXNorm{2}{2}{\Dx\dx(\B v)}\rb.
\end{equation*}

Combining all estimates gives that
\begin{equation}
\Lambda_T^s(\Psi(v))\le C\lb \LTLXNorm{2}{2}{\dx(\B
v)}+\LTLXNorm{2}{2}{\Dx\dx(\B v)} \rb,
\label{Ineq:EstimateOfLambdaOne}
\end{equation}
and since $\Psi(v)-\Psi(\tilde{v})=\Psi(v-\tilde{v})$,
\begin{equation}
\Lambda_T^s(\Psi(v)-\Psi(\tilde{v}))\le C\lb
\LTLXNorm{2}{2}{\dx(\B(v-\tilde{v}))}+\LTLXNorm{2}{2}{\Dx\dx(\B(v-\tilde{v}))}
\rb. \label{Ineq:EstimateOfLambdaTwo}
\end{equation}
Thus, to prove $\Phi+\Psi$ is a strict contraction we need
estimates of the quantities appearing on the right hand side of
\eqref{Ineq:EstimateOfLambdaOne} and
\eqref{Ineq:EstimateOfLambdaTwo}.

H\"{o}lder's inequality gives
\begin{align*}
\LTLXNorm{2}{2}{\dx(\B v)}&\le
\LTLXNorm{2}{\infty}{\B'}\LTLXNorm{\infty}{2}{v}+\LTLXNorm{2}{\infty}{b}\LTLXNorm{\infty}{2}{\dx
v},\\ \LTLXNorm{2}{2}{\Dx\dx(\B v)}&\le
\LTLXNorm{2}{\infty}{\B''}\LTLXNorm{\infty}{2}{v}+\LTLXNorm{2}{\infty}{\B'}\LTLXNorm{\infty}{2}{\dx
v}+\LTLXNorm{2}{2}{\Dx(\B\dx v)}.
\end{align*}
We need to estimate
\begin{equation}
\LTLXNorm{2}{2}{\Dx(\B\dx v)}\le \LTLXNorm{2}{2}{\comm{\Dx}{\B}\dx
v}+\LTLXNorm{2}{2}{\B\Dx\dx v}. \label{Ineq:dxdxbv}
\end{equation}
The first term on the right hand side is bounded using Lemma
\ref{Lemma:CommEstimate}.  We obtain that
\begin{equation*}
\LTLXNorm{2}{2}{\comm{\Dx}{\B}\dx v}\le C \|\widehat{\B'}\|_{L_T^2
L_X^1}\LTLXNorm{\infty}{2}{\dx v},
\end{equation*}
where $\|\widehat{\B'}\|_{L_X^1}$ is the $L^1$ norm of
$\widehat{\B'}$ in the frequency variable.  Using H\"{o}lder's
inequality, the second term in \eqref{Ineq:dxdxbv} is bounded as
\begin{equation*}
\LTLXNorm{2}{2}{\B\Dx\dx v}=\LXLTNorm{2}{2}{\B\Dx\dx
v}\le\LXLTNorm{2}{\infty}{\B}\LXLTNorm{\infty}{2}{\Dx \dx v}.
\end{equation*}
Combining all the estimates implies $\Lambda_T^1(\Psi(v))\le
C\|\B\|_{XT}\Lambda_T^1(v)$ and
$\Lambda^T(\Psi(v)-\Psi(\tilde{v}))\le
C\|\B\|_{XT}\Lambda_T^1(v-\tilde{v})$, where
\begin{equation*}
\|\B\|_{XT,1}:=\LTWXNorm{2}{2}{\infty}{\B}+\LXLTNorm{2}{\infty}{\B}+\|\widehat{\B'}\|_{L_T^2
L_X^1}.
\end{equation*}

If $\varepsilon=C\|\B\|_{XT,1}$, then the above estimates imply
that $\Psi:B_{X_T^s}(a)\rightarrow B_{X_T^s}(\varepsilon a)$ and
\begin{equation}
\Lambda_T^1(\Psi(v)-\Psi(\tilde{v}))\le
\varepsilon\Lambda_T^1(v-\tilde{v}). \label{Ineq:AppendixGWPPsi}
\end{equation}
Thus, $\Phi+\Psi:B_{X_T^s}(a)\rightarrow
B_{X_T^s}(\lb\varepsilon+\epsilon\rb a)$ and
$\Lambda_T^1((\Phi+\Psi)(v)-(\Phi+\Psi)(\tilde{v}))\le
(\varepsilon+\epsilon)\Lambda_T^1(v-\tilde{v})$.  Furthermore, if
we take $\varepsilon+\epsilon<1$, then $\Phi+\Psi$ is a strict
contraction on $B_{X_T^s}(a)$. Invoking the fixed point theorem
completes the proof of the local existence and uniqueness.

Kenig, Ponce and Vega \cite{KePoVe1993} also proved that for all
$0<T_1<T$,
\begin{equation*}
\Lambda_{T_1}^1(\Phi_{v_0}(v)-\Phi_{\tilde{v}_0}(\tilde{v}))\le
C\lb
 \HsNorm{1}{v_0-\tilde{v}_0}+f(T_1)(\Lambda_{T_1}^1(v)+\Lambda_{T_1}^1(\tilde{v}))\Lambda_{T_1}^1(v-\tilde{v})\rb,
\end{equation*}
where $f(T_1)\rightarrow 0$ as $T_1\rightarrow 0$ and $\Phi_{v_0}$
is the map associated to the fixed point problem with initial
condition $v_0$.  The map $\Psi$ is independent of initial
condition; therefore, the triangle inequality and estimate
\eqref{Ineq:AppendixGWPPsi} imply
\begin{equation*}
\Lambda_{T_1}^1((\Phi_{v_0}+\Psi_{v_0})(v)-(\Phi_{\tilde{v}_0}+\Psi_{\tilde{v}_0})(\tilde{v}))\le
C\lb
 \HsNorm{1}{v_0-\tilde{v}_0}+[f(T_1)(\Lambda_{T_1}^1(v)+\Lambda_{T_1}^1(\tilde{v}))+\|b\|_{XT}]\Lambda_{T_1}^1(v-\tilde{v})\rb.
\end{equation*}
Let $v$ be a solution to \eqref{Eqn:AppendixbKdV} with initial
condition $v_0$ and similarly for $\tilde{v}$.  Then, if $T_1$ and
$\|b\|_{XT}$ are small enough, $\Lambda_{T_1}^1(v-\tilde{v})\le
C\HsNorm{1}{v_0-\tilde{v}_0}$.  This proves continuity of the
solution with respect to the initial condition and completes the
proof of local wellposedness of \eqref{Eqn:AppendixbKdV} in
$\Hs{1}$.

To extend local wellposedness to global wellposedness we require
the identities
\begin{equation}
\dt \HamiltonianWithPotential(u)=\frac{1}{2}\intR{(\dt\B)u^2}\
\mbox{and}\ \dt\LpNorm{2}{u}^2=\intR{\B' u^2}
\label{Eqn:AppendixGWPConservationIndents}
\end{equation}
to hold for all $\Hs{1}$ solutions to $\eqref{Eqn:AppendixbKdV}$.
When $u\in \Hs{3}$ both of these follow trivially by integration
by parts.  We appeal to a density argument to prove that the
identities continue to hold in $\Hs{1}$.  Let $\{{u_0}_n\}$ be a
sequence of initial values in $\Hs{3}$ converging in $\Hs{1}$ to
$u_0$.  Then, if \eqref{Eqn:AppendixbKdV} is locally wellposed in
$\Hs{3}$, there are corresponding solutions $u_n\in
C([-T,T],\Hs{3})$ with $u_n(0)={u_0}_n$ and
\begin{equation*}
\lim_{n\rightarrow 0}\sup_{[-T,T]}\HsXNorm{1}{u_n-u}=0.
\end{equation*}
We have used that the time interval appearing in the local
wellposedness result depends continuously only on the $\Hs{1}$
norm of the initial condition.  Hence,
\begin{equation*}
\lim_{n\rightarrow \infty}\dt
\HamiltonianWithPotential(u_n)=\lim_{n\rightarrow\infty}\frac{1}{2}\intR{(\dt\B)u_n^2}=\frac{1}{2}\intR{(\dt\B)u^2}.
\end{equation*}
Since $\dt H(u_n)\rightarrow \dt H(u)$ in distribution, the first
identity of \eqref{Eqn:AppendixGWPConservationIndents} holds.
Similarly,
\begin{equation*}
\lim_{n\rightarrow \infty}\dt\LpXNorm{2}{u_n}^2=\lim_{n\rightarrow
\infty}\intR{\B' u_n^2}=\intR{\B' u^2},
\end{equation*}
and since $\dt\LpXNorm{2}{u_n}^2\rightarrow\dt\LpXNorm{2}{u}^2$ in
distribution, the second identity in
\eqref{Eqn:AppendixGWPConservationIndents} also holds.  The above
assumed local wellposedness in $\Hs{3}$.  The proof of this fact
proceeds as above and one finds that \eqref{Eqn:AppendixbKdV} is
locally wellposed in $\Hs{3}$ if
\begin{equation*}
\|\B\|_{XT,3}:=\LTWXNorm{2}{4}{\infty}{\B}+\LXLTNorm{2}{\infty}{\B}+\|\widehat{\B'}\|_{L^2_T
L_X^1}
\end{equation*}
is small enough.

We now extend the local result to a global result.  The identities
of \eqref{Eqn:AppendixGWPConservationIndents} imply
\begin{equation*}
\dd{t}\LpNorm{2}{u}^2=\intR{\B'u^2}\le \av\ev\LpNorm{2}{u}^2\
\mbox{and}\ \dt
H(u)=\frac{1}{2}\intR{(\dt\B)u^2}\le\frac{\av\ev}{2}\LpNorm{2}{u}^2.
\end{equation*}
Integrating the first by Gronwall's inequality implies
$\LpNorm{2}{u}\le \LpNorm{2}{u_0} \exp(\av\ev t)$.  Substituting
this bound into the above bound on the time derivative of the
Hamiltonian and integrating gives
\begin{equation*}
\frac{1}{2}\LpNorm{2}{\dx u}^2\le
|H(u_0)|+\frac{\tv}{2\ev}\LpNorm{2}{u_0}^2\exp(\av\ev
t)+\LpNorm{2}{u}^2+\frac{1}{2}\left|\intR{\B u^2}\right|.
\end{equation*}
Using the bound on $\LpNorm{2}{u}^2$ again then gives
\begin{equation}
\frac{1}{2}\LpNorm{2}{\dx u}^2\le
|H(u_0)|+\lb1+\frac{\tv}{2\ev}+\frac{1}{2}\av\rb\LpNorm{2}{u_0}^2\exp(\av\ev
t) \label{Ineq:GlobalSolutionThree}
\end{equation}
This inequality implies global existence.  Indeed, say there is a
time $T$ such that $\lim_{t\rightarrow T}\HsNorm{1}{u}=\infty$.
This clearly contradicts \eqref{Ineq:GlobalSolutionThree}.
Uniqueness follows from uniqueness of local solutions.
\end{proof}


\section{Proof of Lemma \ref{Lemma:RegularizedSymplecticOperatorProperties}}
\label{Appendix:PropertiesOfRegSympOpAndRestriction}
\indent\indent Commutativity and the relation
$\dx\RegSympOp=I-\a\RegSympOp$ are direct consequences of
$(\dx+\a)\RegSympOp=I$.  Commutativity with ${\cal S}_a$ is proved
using that
\begin{equation*}
\lb\p_{x-a}+\a\rb^{-1}:g\mapsto e^{-\a(x-a)}\int_{-\infty}^{x-a}
g(y)e^{\a(y-a)}\, d(y-a)
\end{equation*}
and $\dx+\a=\p_{x-a}+\a$.  We prove statements two and five using
the above explicit formula with $a=0$.  Indeed, the inequality
\begin{equation*}
|\RegSympOp\phi|\le \intR{|\phi(x)|}=\LpNorm{1}{\phi}
\end{equation*}
gives statement two, and since $e^{\a(x-y)}-1\le \a
|x-y|e^{\a(x-y)}$, the inequality
\begin{equation*}
\left|\ip{\phi}{\RegSympOp\psi}-\ip{\phi}{\dx^{-1}\psi}\right|\le
\a \intR{|\phi(x)|\int_{-\infty}^\infty|\psi(y)|(|x|+|y|)\,dy}.
\end{equation*}
gives statement five if $x\phi$ and $x\psi$ are integrable.

We prove the remaining statements in Fourier space.  Let
$\hat{\phi}$ be the Fourier transform of $\phi$. Plancherel's
theorem implies
\begin{align}
\LpNorm{2}{\RegSympOp\phi}^2&=\int_{-\infty}^\infty
(k^2+\a^2)^{-1}|\hat{\phi}(k)|^2\, dk\nonumber\\
&=\int_{-\infty}^\infty (k^2+\a^2)^{-1}|\hat{\phi}(0)|^2\,
dk+\int_{-\infty}^\infty
(k^2+\a^2)^{-1}(|\hat{\phi}(k)|^2-|\hat{\phi}(0)|^2)\, dk.
\label{Eqn:ProofRegSympOpPlancherel}
\end{align}
The first equality immediately gives the third statement since
$\LpNorm{\infty}{\hat{\phi}}\le \LpNorm{1}{\phi}$.  A similar
argument gives statement four.  To prove the last statement, we
concentrate on the second integral of
\eqref{Eqn:ProofRegSympOpPlancherel} since the first is easily
computed to be
\begin{equation*}
\frac{\pi}{\a}\lb\intR{\phi}^2\rb.
\end{equation*}
When $\phi$, $x\phi$, and $x^2\phi$ are integrable, $\hat{\phi}$
and $|\hat{\phi}|^2$ are twice differentiable.  Furthermore, since
$|\hat{\phi}|^2$ is even, Taylor's theorem implies
$|\hat{\phi}(k)|^2-|\hat{\phi}(0)|^2=\O{k^2}.$ Thus,
$(k^2+\a^2)^{-1}(|\hat{\phi}(k)|^2-|\hat{\phi}(0)|^2)$ is
integrable for all $\a\in\Rp$ and
\begin{equation*}
\int_{-\infty}^\infty
(k^2+\a^2)^{-1}(|\hat{\phi}(k)|^2-|\hat{\phi}(0)|^2)\, dk=\O{1}.
\end{equation*}
This completes the proof.

\section{Estimates of Nonlinear Remainders}
\indent\indent \label{Appendix:EstimateNonlinearRemainders} Define
\begin{align*}
\NA{\e}:=-\intR{F(\Qca+\e)-F(\Qca)-F'(\Qca)\e-\frac{1}{2}F''(\Qca)\e^2}
\end{align*}
and
\begin{align}
\NpA{\e}:=-\lb f(\Qca+\e)-f(\Qca)-f'(\Qca)\e\rb. \label{Eqn:NpA}
\end{align}
Note that $\NpA{\e}=\p_\xi\NA{\e}$ under the $\Lp{2}$ pairing.
\begin{lemma}
\label{Lemma:NonlinearEstimates} If $\HsNorm{1}{\e}\le 1$ and
$f\in C^k(\R)$ for some $k\ge 3$, with $f^{(k)}\in\Lp{\infty}$,
then there are positive constants $C_1$, $C_2$, and $C_3$ such
that
\begin{enumerate}
    \item $\LpNorm{2}{\NpA{\e}}\le C_1\HsNorm{1}{\e}^2$,\label{Item:NonlinearEstimateNpA}
    \item $\LpNorm{2}{\NpA{\e}+\frac{1}{2}f''(\Qca)\e^2}\le
    C_2\HsNorm{1}{\e}^3$,\label{Item:NonlinearEstimateNpAMore}
    \item $\left|\NA{\e}\right|\le C_3\HsNorm{1}{\e}^3$.\label{Item:NonlinearEstimateNA}
\end{enumerate}
\end{lemma}
\begin{proof}
Taylor's remainder theorem implies
\begin{align*}
\NpA{\e}=-\sum_{n=2}^{k-1}\frac{1}{n!}f^{(n)}(\Qca)\e^n-R(\Qca,
\e),
\end{align*}
where, since $f^{(k)}\in\Lp{\infty}$, $|R(\Qca,\e)|\le C|\e|^k$.
Recall that $\Qca$ is continuous and decays exponentially to zero.
Together with the assumption that $f\in C^k(\R)$, this implies
$f^{(n)}(\Qca)\in\Lp{\infty}$ for $2\le n\le k-1$.  Thus, after
pulling out the largest constant,
\begin{align*}
\LpNorm{2}{\NpA{\e}}\le C\sum_{n=2}^k\LpNorm{2}{\e^n}.
\end{align*}
To obtain statement \ref{Item:NonlinearEstimateNpA} we use the
bound $\LpNorm{2}{\e^n}\le C\HsNorm{1}{\e}^n$, which is obtained
from the inequality $\LpNorm{\infty}{\e}\le C\HsNorm{1}{\e}$ and
the assumption that $\HsNorm{1}{\e}\le 1$.

Clearly, slight modification of the above proof gives items
\ref{Item:NonlinearEstimateNpAMore} and
\ref{Item:NonlinearEstimateNA}.  For the latter we use that the
assumptions on $f$ imply $F\in C^{k+1}(\R)$ with
$F^{(k+1)}\in\Lp{\infty}$.
\end{proof}

\bibliography{Bibliography}

\def\cydot{\leavevmode\raise.4ex\hbox{.}} \def\cprime{$'$}
\begin{thebibliography}{10}

\bibitem{Ar2000}
Enrico Arbarello.
\newblock Sketches of {K}d{V}.
\newblock In {\em Symposium in Honor of C. H. Clemens (Salt Lake City, UT,
  2000)}, volume 312 of {\em Contemp. Math.}, pages 9--69. Amer. Math. Soc.,
  Providence, RI, 2002.

\bibitem{Be1972}
T.~B. Benjamin.
\newblock The stability of solitary waves.
\newblock {\em Proc. Roy. Soc. (London) Ser. A}, 328:153--183, 1972.

\bibitem{BeLi1983}
H.~Berestycki and P.-L. Lions.
\newblock Nonlinear scalar field equations. {I}. {E}xistence of a ground state.
\newblock {\em Arch. Rational Mech. Anal.}, 82(4):313--345, 1983.

\bibitem{Bo1975}
J.~Bona.
\newblock On the stability theory of solitary waves.
\newblock {\em Proc. Roy. Soc. London Ser. A}, 344(1638):363--374, 1975.

\bibitem{BoDo95}
J.~L. Bona, V.~A. Dougalis, O.~A. Karakashian, and W.~R. McKinney.
\newblock Conservative, high-order numerical schemes for the generalized
  {K}orteweg-de {V}ries equation.
\newblock {\em Philos. Trans. Roy. Soc. London Ser. A}, 351(1695):107--164,
  1995.

\bibitem{BoDo96}
J.~L. Bona, V.~A. Dougalis, O.~A. Karakashian, and W.~R. McKinney.
\newblock The effect of dissipation on solutions of the generalized
  {K}orteweg-de {V}ries equation.
\newblock {\em J. Comput. Appl. Math.}, 74(1-2):127--154, 1996.
\newblock TICAM Symposium (Austin, TX, 1995).

\bibitem{BoSm1975}
J.~L. Bona and R.~Smith.
\newblock The initial-value problem for the {K}orteweg-de {V}ries equation.
\newblock {\em Philos. Trans. Roy. Soc. London Ser. A}, 278(1287):555--601,
  1975.

\bibitem{BoSo94}
J.~L. Bona and A.~Soyeur.
\newblock On the stability of solitary-waves solutions of model equations for
  long waves.
\newblock {\em J. Nonlinear Sci.}, 4(5):449--470, 1994.

\bibitem{BoDo86}
Jerry~L. Bona, Vassilios~A. Dougalis, and Ohannes~A. Karakashian.
\newblock Fully discrete {G}alerkin methods for the {K}orteweg-de {V}ries
  equation.
\newblock {\em Comput. Math. Appl. Ser. A}, 12(7):859--884, 1986.

\bibitem{BoDo91}
Jerry~L. Bona, Vassilios~A. Dougalis, Ohannes~A. Karakashian, and William~R.
  McKinney.
\newblock Fully-discrete methods with grid refinement for the generalized
  {K}orteweg-de {V}ries equation.
\newblock In {\em Viscous profiles and numerical methods for shock waves
  (Raleigh, NC, 1990)}, pages 1--11. SIAM, Philadelphia, PA, 1991.

\bibitem{BrJe2000}
J.~C. Bronski and R.~L. Jerrard.
\newblock Soliton dynamics in a potential.
\newblock {\em Math. Res. Lett.}, 7(2-3):329--342, 2000.

\bibitem{BuPe1992}
V.~S. Buslaev and G.~S. Perel{\cprime}man.
\newblock Scattering for the nonlinear {S}chr\"odinger equation: states that
  are close to a soliton.
\newblock {\em Algebra i Analiz}, 4(6):63--102, 1992.

\bibitem{BuSu2003}
Vladimir~S. Buslaev and Catherine Sulem.
\newblock On asymptotic stability of solitary waves for nonlinear
  {S}chr\"odinger equations.
\newblock {\em Ann. Inst. H. Poincar\'e Anal. Non Lin\'eaire}, 20(3):419--475,
  2003.

\bibitem{CoKe2001}
J.~Colliander, M.~Keel, G.~Staffilani, H.~Takaoka, and T.~Tao.
\newblock Global well-posedness for {K}d{V} in {S}obolev spaces of negative
  index.
\newblock {\em Electron. J. Differential Equations}, pages No. 26, 7 pp.
  (electronic), 2001.

\bibitem{CoSt1999}
J.~Colliander, G.~Staffilani, and H.~Takaoka.
\newblock Global wellposedness for {K}d{V} below {$L\sp 2$}.
\newblock {\em Math. Res. Lett.}, 6(5-6):755--778, 1999.

\bibitem{CrGu2004}
W.~Craig, P.~Guyenne, D.P. Nicholls, and C.~Sulem.
\newblock Hamiltonian long wave expansions for water waves over a rough bottom.
\newblock {\em Proceedings of the Royal Society of London A, (to appear)}.

\bibitem{CrGr1994}
Walter Craig and Mark~D. Groves.
\newblock Hamiltonian long-wave approximations to the water-wave problem.
\newblock {\em Wave Motion}, 19(4):367--389, 1994.

\bibitem{CrSu2000}
Walter Craig and Catherine Sulem.
\newblock The water-wave problem and its long-wave and modulational limits.
\newblock In {\em Mathematical and numerical aspects of wave propagation
  (Santiago de Compostela, 2000)}, pages 14--23. SIAM, Philadelphia, PA, 2000.

\bibitem{DeZh93}
P.~Deift and X.~Zhou.
\newblock A steepest descent method for oscillatory {R}iemann-{H}ilbert
  problems. {A}symptotics for the {MK}d{V} equation.
\newblock {\em Ann. of Math. (2)}, 137(2):295--368, 1993.

\bibitem{DeJo2004}
S.~I. Dejak and B.~L.~G. Jonsson.
\newblock Long time dynamics of {mKdV} solitons (in preparation 2004).

\bibitem{FrGu2003}
J.~Fr\"{o}hlich, S.~Gustafson, B.L.G Jonsson, and I.M. Sigal.
\newblock Solitary wave dynamics in an external potential (in-print).
\newblock {\em Comm. Math. Phys.}, 2003.

\bibitem{FrTs2003}
J\"{u}rg Fro\"{o}hlich, Tai-Peng Tsai, and Horng-Tzer Yau.
\newblock On the point particle ({N}ewtonian) limit of the non-linear {H}artree
  equation.
\newblock {\em Comm. Math. Phys.}, 225(2):223--274, 2002.

\bibitem{GaSi2004}
Zhou Gang and I.M. Sigal.
\newblock Asymptotic stability of nonlinear {Sch\"{o}dinger} equation with
  potential ({Preprint, Toronto} 2004).

\bibitem{GrShSt87}
Manoussos Grillakis, Jalal Shatah, and Walter Strauss.
\newblock Stability theory of solitary waves in the presence of symmetry. {I}.
\newblock {\em J. Funct. Anal.}, 74(1):160--197, 1987.

\bibitem{GuSi2003}
S.~Gustafson and Sigal I.M.
\newblock {\em Mathematical Concepts of Quantum Mechanics}.
\newblock Springer-Verlag, New York, 2003.

\bibitem{JeKa1970}
Alan Jeffrey and Tsunehiko Kakutani.
\newblock Stability of the {B}urgers shock wave and the {K}orteweg-de {V}ries
  soliton.
\newblock {\em Indiana Univ. Math. J.}, 20:463--468, 1970/1971.

\bibitem{Jo1973}
R.~S. Johnson.
\newblock On the development of a solitary wave moving over an uneven bottom.
\newblock {\em Proc. Cambridge Philos. Soc.}, 73:183--203, 1973.

\bibitem{Ka1983}
Tosio Kato.
\newblock On the {C}auchy problem for the (generalized) {K}orteweg-de {V}ries
  equation.
\newblock In {\em Studies in applied mathematics}, volume~8 of {\em Adv. Math.
  Suppl. Stud.}, pages 93--128. Academic Press, New York, 1983.

\bibitem{KePoVe1993}
Carlos~E. Kenig, Gustavo Ponce, and Luis Vega.
\newblock Well-posedness and scattering results for the generalized
  {K}orteweg-de {V}ries equation via the contraction principle.
\newblock {\em Comm. Pure Appl. Math.}, 46(4):527--620, 1993.

\bibitem{KePoVe1996}
Carlos~E. Kenig, Gustavo Ponce, and Luis Vega.
\newblock A bilinear estimate with applications to the {K}d{V} equation.
\newblock {\em J. Amer. Math. Soc.}, 9(2):573--603, 1996.

\bibitem{Ke02}
Sahbi Keraani.
\newblock Semiclassical limit of a class of {S}chr\"odinger equations with
  potential.
\newblock {\em Comm. Partial Differential Equations}, 27(3-4):693--704, 2002.

\bibitem{KoDe1895}
D.J. Korteweg and F.~{de Vries}.
\newblock On the change of form of long waves advancing in a rectangular canal,
  and on a new type of long stationary waves.
\newblock {\em Philos. Mag.}, 39:422--443, 1895.

\bibitem{Mi1979}
John~W. Miles.
\newblock On the {K}orteweg-de\thinspace {V}ries equation for a gradually
  varying channel.
\newblock {\em J. Fluid Mech.}, 91(1):181--190, 1979.

\bibitem{ReSiI}
Simon Reed and Barry Simon.
\newblock {\em Methods of Modern Mathematical Physics, I. Functional Analysis}.
\newblock Academic Press, San Diego, second edition, 1980.

\bibitem{ReSiIV}
Simon Reed and Barry Simon.
\newblock {\em Methods of Modern Mathematical Physics, IV. Analysis of
  Operators}.
\newblock Academic Press, San Diego, 1980.

\bibitem{RoScSoPreprintII}
I.~Rodnianski, W.~Schlag, and Soffer A.
\newblock Asymptotic stability of n-soliton state of nls, arxiv:math.ap.

\bibitem{RoScSoPreprint}
I.~Rodnianski, W.~Schlag, and Soffer A.
\newblock Dispersive analysis of charge transfer models, arxiv:math.ap.

\bibitem{ScWa2000}
Guido Schneider and C.~Eugene Wayne.
\newblock The long-wave limit for the water wave problem. {I}. {T}he case of
  zero surface tension.
\newblock {\em Comm. Pure Appl. Math.}, 53(12):1475--1535, 2000.

\bibitem{SoWe90}
A.~Soffer and M.~I. Weinstein.
\newblock Multichannel nonlinear scattering for nonintegrable equations.
\newblock {\em Comm. Math. Phys.}, 133(1):119--146, 1990.

\bibitem{St1977}
Walter~A. Strauss.
\newblock Existence of solitary waves in higher dimensions.
\newblock {\em Comm. Math. Phys.}, 55(2):149--162, 1977.

\bibitem{TsYa2002III}
Tai-Peng Tsai and Horng-Tzer Yau.
\newblock Asymptotic dynamics of nonlinear {S}chr\"odinger equations:
  resonance-dominated and dispersion-dominated solutions.
\newblock {\em Comm. Pure Appl. Math.}, 55(2):153--216, 2002.

\bibitem{TsYa2002II}
Tai-Peng Tsai and Horng-Tzer Yau.
\newblock Relaxation of excited states in nonlinear {S}chr\"odinger equations.
\newblock {\em Int. Math. Res. Not.}, (31):1629--1673, 2002.

\bibitem{TsYa2002I}
Tai-Peng Tsai and Horng-Tzer Yau.
\newblock Stable directions for excited states of nonlinear {S}chr\"odinger
  equations.
\newblock {\em Comm. Partial Differential Equations}, 27(11-12):2363--2402,
  2002.

\bibitem{GrPu1993}
E.~van Groesen and S.~R. Pudjaprasetya.
\newblock Uni-directional waves over slowly varying bottom. {I}. {D}erivation
  of a {K}d{V}-type of equation.
\newblock {\em Wave Motion}, 18(4):345--370, 1993.

\bibitem{We1985}
M.I. Weinstein.
\newblock Modulational stability of ground states of nonlinear
  {{Schr\"{o}dinger}} equations.
\newblock {\em SIAM J. Math. Anal.}, 16(3):472--491, 1985.

\bibitem{Wi1974}
G.B. Whitham.
\newblock {\em Linear and Nonlinear Waves}.
\newblock Wiley, New York, 1974.

\bibitem{YoLi1994}
Sung~B. Yoon and Philip L.-F. Liu.
\newblock A note on {H}amiltonian for long water waves in varying depth.
\newblock {\em Wave Motion}, 20(4):359--370, 1994.

\end{thebibliography}
\bibliographystyle{plain}

\end{document}